\documentclass[apj,twocolumn]{emulateapj}

\usepackage{natbib}
\usepackage{appendix}
\usepackage{amsmath}
\usepackage{amssymb}
\usepackage{graphicx}
\usepackage{verbatim}
\usepackage{lscape}

\slugcomment{Submitted to ApJ on Mar 20th 2014}

\shorttitle{High Redshift Clusters in the {\it Spitzer} SPT Deep Field Survey}
\shortauthors{Rettura et al.}

\begin{document}

%% LaTeX will automatically break titles if they run longer than
%% one line. However, you may use \\ to force a line break if
%% you desire.

\title{Candidate Clusters of Galaxies at $z>1.3$ Identified in the {\it Spitzer} SPT Deep Field Survey}

%% Use \author, \affil, and the \and command to format
%% author and affiliation information.
%% Note that \email has replaced the old \authoremail command
%% from AASTeX v4.0. You can use \email to mark an email address
%% anywhere in the paper, not just in the front matter.
%% As in the title, use \\ to force line breaks.

\author{A. Rettura\altaffilmark{1,2}, J. Martinez-Manso\altaffilmark{3},  D. Stern\altaffilmark{1}, S. Mei\altaffilmark{4,5,6}, M.L.N. Ashby\altaffilmark{7}, M. Brodwin\altaffilmark{8}, D. Gettings\altaffilmark{3}, A.H. Gonzalez\altaffilmark{3}, S.A. Stanford\altaffilmark{9,10} and J.G. Bartlett\altaffilmark{11,1}}

\altaffiltext{1}{Jet Propulsion Laboratory, California Institute of Technology, MS 169-234, Pasadena, CA 91109, USA}
\altaffiltext{2}{Department of Astronomy, California Institute of Technology, MS 249-17, Pasadena, CA 91125, USA}
\altaffiltext{3}{Department of Astronomy, University of Florida, Gainesville, FL 32611, USA}
\altaffiltext{4}{GEPI, Observatoire de Paris, Section de Meudon, Meudon Cedex, France}
\altaffiltext{5}{University of Paris Denis Diderot, 75205 Paris Cedex 13, France}
\altaffiltext{6}{Infrared Processing and Analysis Center, Pasadena, CA 91125, USA}
\altaffiltext{7}{Harvard-Smithsonian Center for Astrophysics, Cambridge, MA 02138, USA}
\altaffiltext{8}{Department of Physics and Astronomy, University of Missouri, Kansas City, MO 64110, USA}
\altaffiltext{9}{Department of Physics, University of California, Davis, CA 95616, USA}
\altaffiltext{10}{Institute of Geophysics and Planetary Physics, Lawrence Livermore National Laboratory, Livermore, CA 94551, USA}
\altaffiltext{11}{APC, AstroParticule et Cosmologie, Universite Paris Diderot, CNRS/IN2P3, CEA/lrfu, Observatoire de Paris, Sorbonne Paris Cite, 75205 Paris Cedex 13, France}

%% Notice that each of these authors has alternate affiliations, which
%% are identified by the \altaffilmark after each name.  Specify alternate
%% affiliation information with \altaffiltext, with one command per each
%% affiliation.

%\altaffiltext{1}{Visiting Astronomer, Cerro Tololo Inter-American Observatory.
%CTIO is operated by AURA, Inc.\ under contract to the National Science
%Foundation.}
%\altaffiltext{2}{Society of Fellows, Harvard University.}
%\altaffiltext{3}{present address: Center for Astrophysics,
%    60 Garden Street, Cambridge, MA 02138}
%\altaffiltext{4}{Visiting Programmer, Space Telescope Science Institute}
%\altaffiltext{5}{Patron, Alonso's Bar and Grill}

%% Mark off your abstract in the ``abstract'' environment. In the manuscript
%% style, abstract will output a Received/Accepted line after the
%% title and affiliation information. No date will appear since the author
%% does not have this information. The dates will be filled in by the
%% editorial office after submission.

\begin{abstract}

 We present 279 galaxy cluster candidates at $z > 1.3$ selected from the 94 deg$^{2}$ {\it Spitzer} South Pole Telescope Deep Field (SSDF) survey. We use a simple %three filter 
algorithm to select candidate high-redshift clusters of galaxies based on  {\it Spitzer}/IRAC mid-infrared data combined with shallow all-sky optical data. 
We identify distant cluster candidates in SSDF adopting an overdensity threshold that results in a high purity (80\%) cluster sample based on tests in the {\it Spitzer} Deep, Wide-Field Survey of the Bo\"otes field. Our simple algorithm detects all three $1.4 < z \leq 1.75$  X-ray detected clusters in  the Bo{\"o}tes field. % IDCS/SDWFS (Ashby et al. 2009) survey. 
The uniqueness of the SSDF survey resides not just in its area, one of the largest contiguous extragalactic fields observed with {\it Spitzer}, but also in its deep, multi-wavelength  coverage by the South Pole Telescope  (SPT), %(George et al., 2012), far-infrared observations from
{\it Herschel}/SPIRE and {\it XMM-Newton}. % XXL Survey (Pierre et al. 2012). 
This rich dataset will allow direct or stacked measurements of  Sunyaev-Zel'dovich effect decrements or X-ray masses for many of the SSDF clusters presented here, and enable systematic study of the most distant clusters on an unprecedented scale.
We measure the angular correlation function of our sample and find that these candidates show strong clustering. Employing the COSMOS/UltraVista photometric catalog in order to infer the redshift distribution of our cluster selection, we find that these clusters have a comoving number density $n_c  = (0.7^{+6.3}_{-0.6}) \times 10^{-7} h^{3} \mathrm{Mpc}^{-3}$%$n_c  = (7.6 \pm 0.6 ) \times 10^{-7} h^{3} \mathrm{Mpc}^{-3}$
 and a spatial clustering correlation scale length $r_0 = (32 \pm 7) h^{-1} \rm{Mpc}$%$r_0 = (24.5 \pm 0.4) h^{-1} \rm{Mpc}$
. Assuming our sample is comprised of dark matter halos above a characteristic minimum mass, $M_{{\rm min}}$, we derive that at $z=1.5$ these clusters reside in halos larger than $M_{{\rm min}} = 1.5^{+0.9}_{-0.7} \times 10^{14} h^{-1} M_{\odot}$ %(9.2 \pm 0.2) \times 10^{13}h^{-1} M_\odot$
. We find the mean mass of our cluster sample to be equal to  $M_{{\rm mean}} = 1.9^{+1.0}_{-0.8} \times 10^{14} h^{-1} M_{\odot}$, thus our sample contains the progenitors of present-day massive galaxy clusters. 
\end{abstract}

%% Keywords should appear after the \end{abstract} command. The uncommented
%% example has been keyed in ApJ style. See the instructions to authors
%% for the journal to which you are submitting your paper to determine
%% what keyword punctuation is appropriate.

%\keywords{galaxies:    galaxies:   high-redshift   ---   galaxies:
%  fundamental  parameters   ---  galaxies:  evolution   ---  galaxies:
%  formation --- galaxies: elliptical --- cosmology: observations}

%galaxies:  fundamental  parameters   ---  galaxies:  evolution 

%% From the front matter, we move on to the body of the paper.
%% In the first two sections, notice the use of the natbib \citep
%% and \citet commands to identify citations.  The citations are
%% tied to the reference list via symbolic KEYs. The KEY corresponds
%% to the KEY in the \bibitem in the reference list below. We have
%% chosen the first three characters of the first author's name plus
%% the last two numeral of the year of publication as our KEY for
%% each reference.

%% Authors who wish to have the most important objects in their paper
%% linked in the electronic edition to a data center may do so by tagging
%% their objects with \objectname{} or \object{}.  Each macro takes the
%% object name as its required argument. The optional, square-bracket 
%% argument should be used in cases where the data center identification
%% differs from what is to be printed in the paper.  The text appearing 
%% in curly braces is what will appear in print in the published paper. 
%% If the object name is recognized by the data centers, it will be linked
%% in the electronic edition to the object data available at the data centers  
%%%% Note that for sources with brackets in their names, e.g. [WEG2004] 14h-090,
%% the brackets must be escaped with backslashes when used in the first
%% square-bracket argument, for instance, \object[\[WEG2004\] 14h-090]{90}).
%%  Otherwise, LaTeX will issue an error. 

\section{Introduction}

Emerging from the cosmic web, galaxy clusters are the most massive
gravitationally bound structures in the universe. Thought to have
begun their assembly at $z > 2$, clusters provide insights into
the growth of large-scale structure as well as the physics that
drives galaxy evolution.  Understanding how and when the most massive
galaxies assemble their stellar mass, stop forming stars, and acquire
their observed morphologies remain outstanding questions.  The
redshift range $1.4 < z < 2$ is a key epoch in this respect:
elliptical galaxies start to become the dominant population in
cluster cores, and star formation in spiral galaxies is being
quenched \citep[e.g.,][]{Blakeslee06, Rosati09, Mei09, Overzier09, Rettura10, Rettura11, Raichoor11, Strazzullo10, Stanford12, Snyder12, Zeimann12, Mei12, Nantais13}.  
Interestingly, some field galaxy studies find that the star formation rate (SFR)-density relation
reverses  at $z=1$ relative  to $z=0$  \citep{Cooper07, Elbaz07}, such that star formation no longer decreases with increasing galaxy density at $z=1$. However some other studies disagree with this result \citep{Patel09, Muzzin12, Scoville13} and conclude the reversal must happen at $z>1$ as they find the local density correlations to be  already in place by  $z = 1$. 
There is also observational evidence for a
progressive increase in the amount of star formation that occurs
in galaxy cluster cores at $z \gtrsim 1.4$ \citep[e.g.,][]{Hilton09, Hayashi10, Tran10, Fassbender11, Tadaki12, Brodwin13, Alberts14}. This suggests  that  significant  star formation  is
occurring  in  high-density  environments at  early  epochs.
 Therefore, increasing evidence points to clusters at $1.5 < z
< 2$ as being the ideal laboratories to study cluster formation and to catch in the act transformations
in their stellar populations.

Until recently, however, this redshift range was essentially unreachable with available instrumentation, %, and called a `desert'.
with clusters at these redshifts  exceedingly challenging to identify from either ground-based optical/near-infrared (NIR) imaging or from X-ray surveys.
Mid-infrared (MIR) imaging with {\it Spitzer} has changed the
 landscape. %Few massive clusters have been confirmed at these redshifts.  
Previous {\it Spitzer} wide-area surveys have proven effective
 at identifying samples of galaxy clusters down to low masses at $1 \lesssim z < 2$ \citep[e.g., SDWFS, SWIRE, CARLA][]{Eisenhardt08, Papovich08, Wilson09, Demarco10, Galametz10, Stanford12, Zeimann12, Brodwin13, Galametz13, Muzzin13, Wylezalek13} where current X-ray observations are restricted to only the most massive systems.
X-ray follow-up has verified several of these MIR-selected clusters, implying masses of a few $10^{14} M_{\odot}$ \citep{Papovich10, Brodwin11, Muzzin13}. To date, however, only a few
clusters have been confirmed at $z > 1.5$, in part due to a lack of sufficiently large {\it Spitzer} surveys.

With the {\it Spitzer}-South Pole Telescope Deep Field survey \citep[SSDF;][]{Ashby13}, we aim to discover hundreds of cluster candidates at these redshifts.  The uniqueness of the SSDF survey resides not just in its area, 94 deg$^2$, one of the largest contiguous extragalactic fields surveyed with {\it Spitzer}, but also in its coverage by deep observations for the Sunyaev-Zel'dovich (SZ) effect by the South Pole Telescope (SPT), with even deeper observations being taken with the new SPT camera, SPTpol \citep{George12}. Approximately one fourth of the SSDF field also has deep X-ray observations from the {\it XMM-Newton} XXL Survey \citep{Pierre11}. This rich multi-wavelength dataset will allow us to determine cluster masses for many of the SSDF clusters at $1 .5< z < 2$, enabling systematic study of the cluster population at an important cosmic epoch.

The structure  of this  paper is as  follows.  The description  of our
datasets comprise \S 2. In \S 3
we describe the  method we employ to identify distant galaxy clusters, and we estimate our sample purity based on analysis of the Bo{\"o}tes field.  
In \S 4 we study the clustering of our sample, deriving the characteristic minimum mass, $M_{\rm min}$, of the dark matter halos in which our clusters reside.  Section 5 summarizes the conclusions of our study. Throughout, we assume a $\Omega_{\Lambda} = 0.73$, $\Omega_{m} = 0.27$ and $H_{0} = 71\ \rm{km} \rm{s}^{-1} \rm{Mpc}^{-1}$ cosmology \citep{Spergel03}, and use magnitudes in the AB system.

\section{The {\it Spitzer} South Pole Telescope Deep Field Survey}

The SSDF, centered at $23^h30^m$, $-55^d00^m$ (J2000), is a wide-area
survey using the {\it Spitzer} Infrared Array Camera
\citep[IRAC;][]{Fazio04} to cover 94 deg$^2$ of extragalactic sky.
We discuss the IRAC and publicly available optical data next.
\citet{Ashby13} summarizes other available data in this field,
including X-ray observations from {\it XMM-Newton} \citep{Pierre11},
shallow near-infrared data from the VISTA Hemisphere Survey (VHS), far-infrared
data from {\it Herschel}/SPIRE \citep{Holder13}, and millimeter data from the SPT \citep{Carlstrom11, Austermann12, Story13}.

\subsection{{\it Spitzer}/IRAC Data}

The SSDF, a post-cryogenic {\it Spitzer} Exploration Science
program, obtained 120s-depth observations in the 3.6$\mu$m and
4.5$\mu$m IRAC bandpasses (hereafter, [3.6], [4.5]).  \citet{Ashby13}
provide detailed information on the survey design, observations,
processing, source extraction, and publicly available data products.
Our study is based on the 4.5 $\mu$m-selected {\it Spitzer}/IRAC
band-merged ([3.6], [4.5]) catalog, which contains $\sim 3.7$ million
distinct sources down to the SSDF $5\sigma$ sensitivity limit of
21.46 AB mag (9.4 $\mu$Jy) at 4.5 $\mu$m; the corresponding $5\sigma$
sensitivity of the 3.6 $\mu$m bandpass is 21.79 AB mag (7.0 $\mu$Jy).
Throughout this paper we use aperture-corrected, 4\arcsec-diameter
aperture magnitudes for the IRAC data.

\subsection{SuperCOSMOS Optical Data}

The SuperCOSMOS survey \citep{Hambly01} provides all-sky optical photometry based on scans of photographic Schmidt survey plates from the UK Schmidt Telescope (UKST) Southern Surveys \citep{Hartley81, Cannon84} and Palomar Oschin Schmidt Telescope Surveys \citep[POSS;][]{Minkowski63, Reid91}. These shallow data provide $I$-band magnitudes down to $I \sim 20.45$ mag (AB) in the SSDF \citep{Hambly01}.  Portions of the SSDF field have deeper optical data from more recent surveys, such as the Blanco Cosmology Survey which reaches $\sim 1 \mu$Jy depths in $griz$ \citep{Desai12}, and the whole field will be covered by the Dark Energy Survey \citep{Mohr12}. However, in the interest of uniformity in our cluster search over the widest possible area, we only consider the relatively shallow SuperCOSMOS optical data in the following analysis.

\begin{figure}
\epsscale{1.0}
\includegraphics[scale=.55,angle=-90]{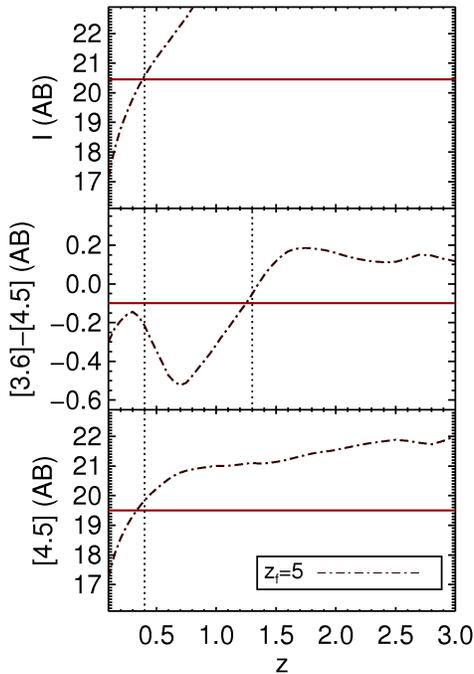}
\caption{The evolution with redshift of SuperCOSMOS $I$-band magnitude (top panel),  $[3.6]-[4.5]$ IRAC color (middle panel) and $[4.5]$ IRAC magnitude (bottom panel) for \citet{BC03} simple stellar population models formed at formation redshifts, $z_{f}$, of 5 (black dot-dashed line).% and 10 (red dashed line).  and normalized to a typical $K=19.8$ mag (Vega) at $z=1.5$.
The IRAC color selection criterion we adopt is optimized  to find galaxies with $z > 1.3$. The $[3.6]-[4.5]$ color serves as a good redshift indicator for $0.7 \lesssim z \lesssim 1.5$; above this redshift, the $[3.6]-[4.5]$ color evolution with redshift flattens. To alleviate a known contamination from foreground interlopers at $z\sim 0.3$ \citep[e.g.,][]{Papovich08, Muzzin13} we also apply two magnitude cuts ($[4.5]>19.5$ and $I > 20.45$). 
}
\label{colorselection}
\end{figure}

\begin{figure}
\epsscale{1.0}
\includegraphics[scale=.40,angle=0]{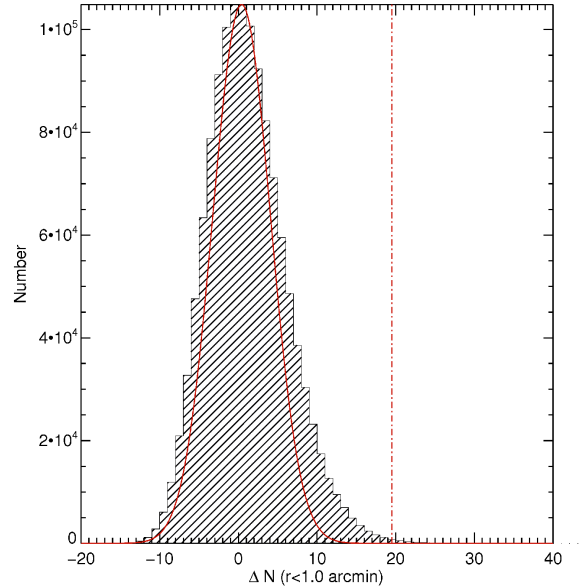}
\caption{Distribution of the completeness-corrected excess number of objects with respect to the local background, $\Delta N$, with $[3.6] - [4.5]>-0.1$,  $19.5<[4.5]<21.46$ and $I>20.45$,  within $1.0\arcmin$ radius from each individual source in the SSDF catalog fulfilling the aforementioned color criteria. The mean number of excess objects is $\langle \Delta N \rangle = 0.4 \pm 3.7$. The red curve shows a Gaussian fit (iteratively clipping at 2$\sigma$). The dashed line indicates the minimum detection significance threshold adopted here of 5.2 times the standard deviations of the Gaussian distribution, corresponding to objects with more than 19.5 similarly selected (completeness corrected) companions within $1.0\arcmin$.}
\label{hist_ssdf}
\end{figure}

\begin{figure}
\epsscale{1.0}
\includegraphics[scale=.23]{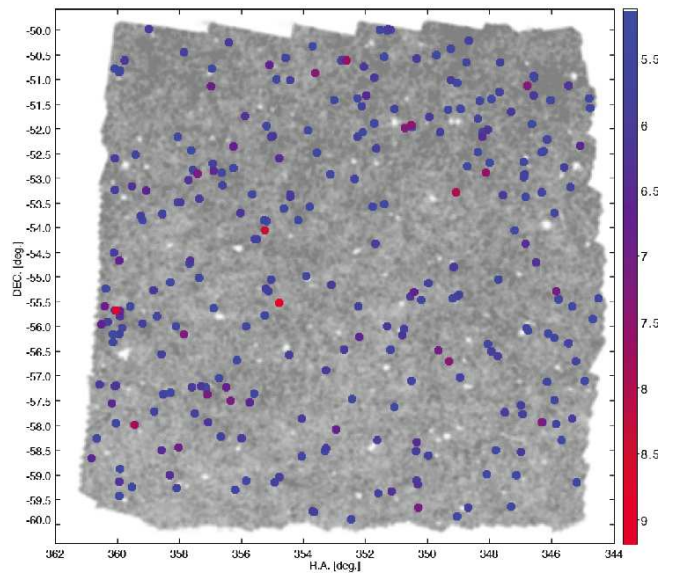}
\caption{Spatial distribution of 279 galaxy cluster candidates with $X_{f} \geq 5.2$ detection significance in the SSDF. Each data point corresponds to one cluster candidate defined in \S 3.1, color-coded by detection significance. The grey background shows the positions of all galaxies in the SSDF field that meet the color criteria described in the text. Note that the white gaps in the backgound map are due to masked areas surrounding bright stars.}
\label{clus_map}
\end{figure}

\section{Identifying High-Redshift Galaxy Clusters}

\subsection{Methodology}

We select  candidate distant galaxy clusters based on their $[3.6]-[4.5]$ galaxy color, following the approach of \citet{Papovich08}; that methodology was proven effective  by discovering a $z = 1.62$ galaxy cluster in the SWIRE-{\it XMM} field \citep{Lonsdale03} using data of similar depth to what is available in the SSDF \citep{Papovich10}. The method takes advantage of the fact that $[3.6]-[4.5]$  color is a linear function of redshift  between $0.7 \lesssim z  \lesssim 1.5$ (Fig. \ref{colorselection}), and thus can be used as an effective redshift indicator. At  $z \gtrsim 1.5$ the color reaches a plateau out to $z\sim 3$. The basis of the selection is that galaxy stellar populations with ages $> 10$ Myr have a prominent bump at $\sim 1.6 \mu$m due to a minimum in the opacity of the H$^{-}$ ion present in the atmospheres of cool stars. This feature is seen in the spectral energy distribution (SED) of essentially all galaxies,  largely independent of star formation history or age \citep{John88, Simpson99, Sawicki02,Sorba10}.
However, above redshift $z \sim 1.5$  the color is not precise enough to be useful to estimate  redshifts other than to constrain the redshift to be larger than $z \gtrsim 1.5$. 

As shown in the middle panel of Fig. \ref{colorselection}, while an IRAC color cut $[3.6]-[4.5]>-0.1$ is effective at distinguishing galaxies at $z>1.3$,
having at least one relatively shallow optical band is very useful for alleviating contamination from foreground interlopers at $z\sim 0.3$ \citep[see discussion in][]{Muzzin13}. 
To this aim, we  apply magnitude cuts in the $I$ and  $[4.5]$ bands to remove most  $z < 0.4$ galaxies (see bottom and top panel of Fig. \ref{colorselection}). Combined, these cuts effectively remove the bulk of the foreground galaxy population at $z < 1.3$. The remaining sources consist predominantly of high-redshift galaxies. We note that other sources of contamination are cool brown dwarfs \citep{Stern07} and powerful AGNs at all redshifts \citep{Stern05}; however, these are expected not to be contaminants \citep[see discussion in][]{Galametz12}.

Therefore, to discover distant clusters of galaxies at $z>1.3$ we have implemented a simple
three-filter algorithm to search for overdensities of galaxies based upon their {\it Spitzer} IRAC color ($[3.6]-[4.5]>-0.1$), their $4.5 \mu$m magnitude   ($[4.5]>19.5$)  and
requiring non-detection in the SuperCOSMOS $I$-band data ($I > 20.45$). Similar algorithms have been demonstrated
to be effective by several programs \citep{Papovich10, Galametz10, Galametz13, Gettings12, Muzzin13}. 

To identify candidate galaxy clusters, for each galaxy in the SSDF catalog that meets the aforementioned criteria we count the number of similarly selected companions within $1.0\arcmin$-radius cells. We correct these counts in cells for completeness using the values in Table 5 of \citet{Ashby13}, derived from a Monte Carlo analysis. When creating an IRAC mosaic, the mapping strategy adopted results in some variations in depth across a large field. Therefore, we search for cells  with the highest significance overdensity of IRAC-selected sources above the local background. To this aim, for each selected object in the catalog, we also count the completeness-corrected number of similarly selected companions in an outer annulus defined by radii of $5\arcmin$ and $8\arcmin$.
After taking into account the difference in surface areas, %per each individual selected source in the SSDF catalog 
we then derive the excess number of selected companions with respect to the local background, $\Delta N$. 

Fig.~\ref{hist_ssdf} shows the distribution of completeness-corrected excess number of companions, $\Delta N$, within $1.0\arcmin$ radius. The mean number of excess galaxies is $\langle \Delta N \rangle = 0.4$. The distribution is skewed toward objects with higher-than-average numbers of companions, suggestive of strong clustering. 
%As pointed out by \citet{Papovich08} if the overdensity of objects results from the projection effect of unassociated objects along the line of sight, or from other random processes, then the distribution in Figure \ref{hist_ssdf} should be consistent with a Gaussian distribution.
Similarly to \citet{Papovich08}, we fit a Gaussian to the distribution iteratively clipping at 2$\sigma$. The best fit is shown in Fig.~\ref{hist_ssdf} and it matches well the low-excess half of the distribution. We find the width of the fitted Gaussian to be $\sigma_{\Delta N} = 3.7$.

The minimum detection significance to adopt, $X_{f_{\rm min}}$, is somewhat arbitrary, trading between completeness and reliability in the derived candidate cluster catalog.
We based our conservative choice on extensive tests performed on similar-depth IRAC and SuperCOSMOS data in the Bo{\"o}tes field,  where ancillary spectroscopic and much deeper multi-wavelength photometric coverage is also present. We describe these tests in section \S 3.2, where we estimate that the threshold adopted here ensures a purity of $\sim 80\%$.
Thus we define our sample of candidate clusters of galaxies to be those with $ \Delta N \geq \langle \Delta N \rangle + X_{f_{\rm min}} \cdot \sigma_{\Delta N} $, 
with $X_{f_{\rm min}} = 5.2$. For the SSDF field this corresponds to objects with $\Delta N \geq 19.5$ excess companions with respect to the local background (indicated by the vertical dot-dashed red line of Fig. \ref{hist_ssdf}). 

By design, many of the objects in overdense regions will be identified as companions to multiple sources by our algorithm. To address this, we adopt a similar approach as done in \citet{Papovich08}. We merge cluster candidates applying a friends-of-friends algorithm with a linking length of $1.0\arcmin$.
This results in a final catalog containing $N_{\rm obs}$=279 candidate clusters; their spatial distribution and color-coded detection significance are shown in Fig. \ref{clus_map}. We note that none of our clusters is found in the list of SPT SZ-detected clusters reported in \citet{Reichardt13}. 
This is not unexpected, since the two samples do not overlap in mass-redshift space: the highest-redshift cluster in Reichardt et al. (2013) is at  z=1.075, below the range to which our cluster finding algorithm is sensitive, while the typical mass of the clusters we find is below the mass threshold of the \citet{Reichardt13} catalog.

\begin{figure*}
\epsscale{1.0}
\includegraphics[scale=1.19]{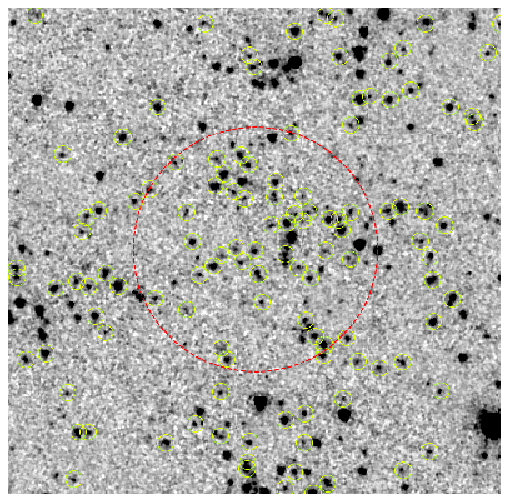}
\includegraphics[scale=1.19]{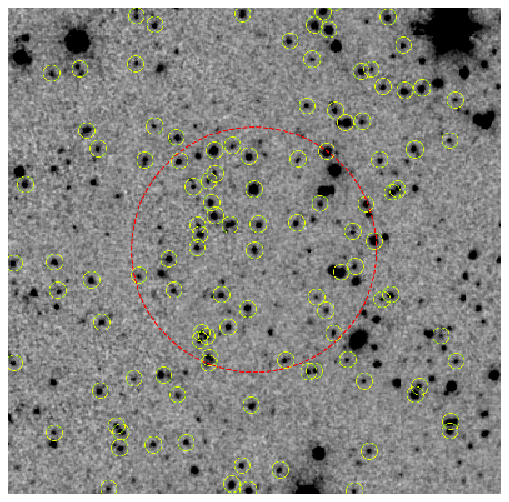}
\includegraphics[scale=1.19]{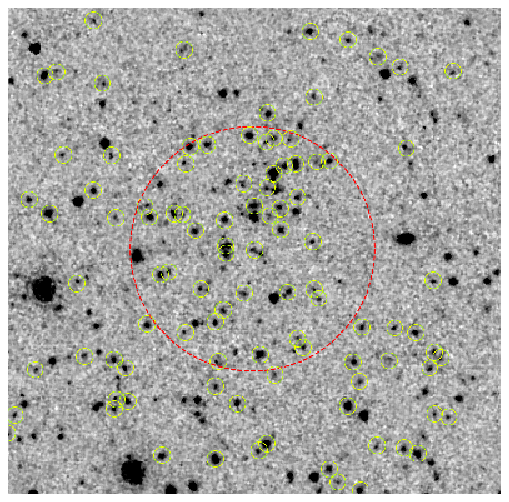}
\includegraphics[scale=1.19]{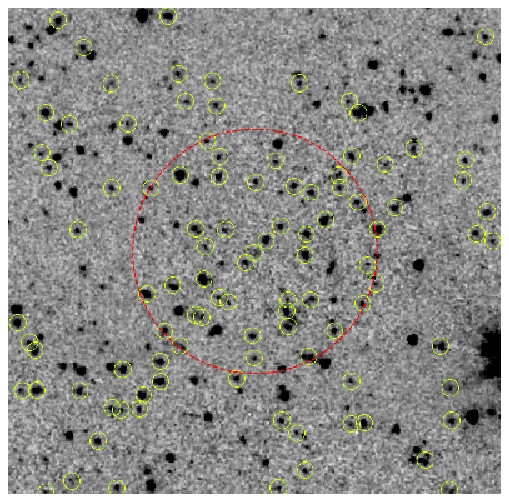}
\includegraphics[scale=1.19]{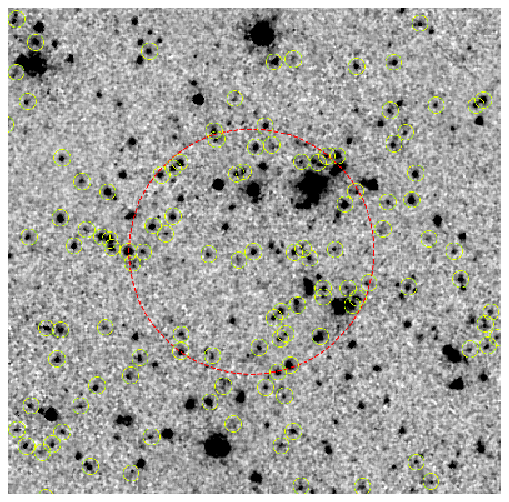}
\includegraphics[scale=1.19]{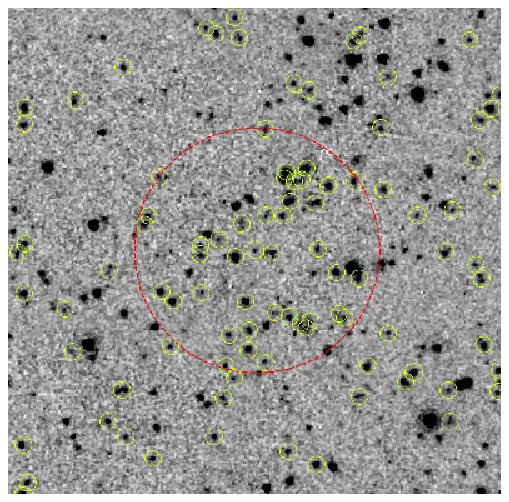}
\includegraphics[scale=1.19]{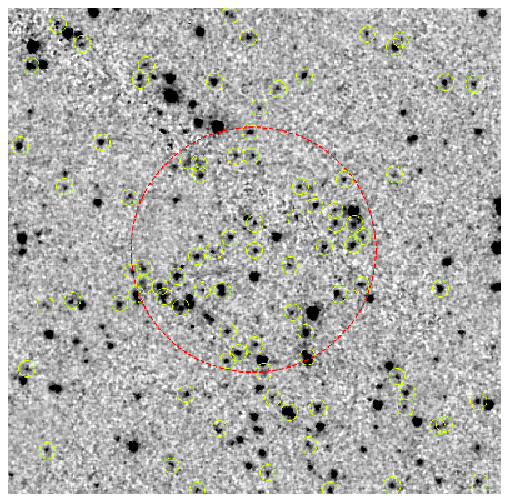}
\includegraphics[scale=1.19]{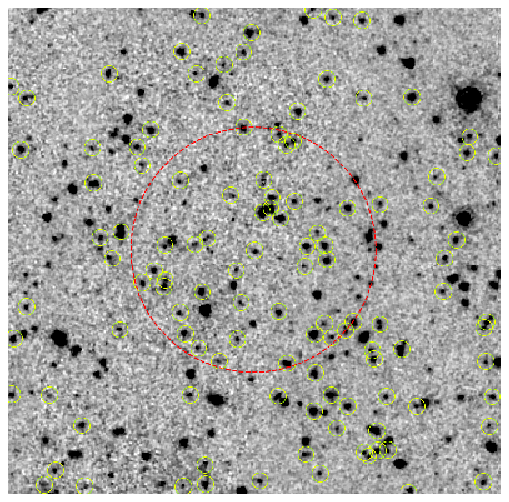}
\includegraphics[scale=1.19]{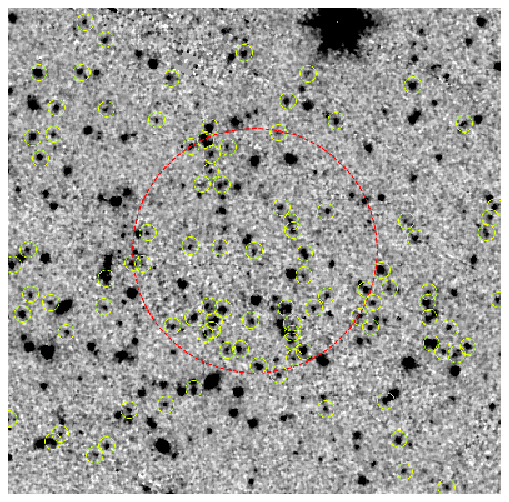}
\includegraphics[scale=1.19]{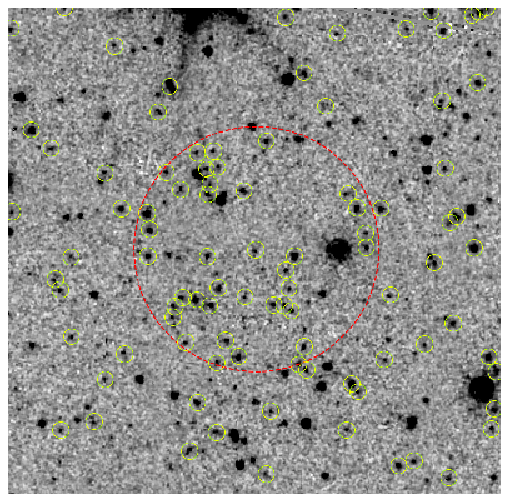}
\caption{{\it Spitzer}/IRAC 4.5$\mu$m images $(4' \times 4')$ of the 10 most significant high-redshift candidate galaxy clusters in the SSDF survey. The yellow circles indicate candidate cluster members. The red circle has a radius of $1.0\arcmin$ that corresponds to an angular diameter distance of $\sim 0.5$ Mpc at $z=1.5$.}
\label{clus_images}
\end{figure*}

\begin{figure*}
\epsscale{1.0}
\includegraphics[scale=.23,angle=-90]{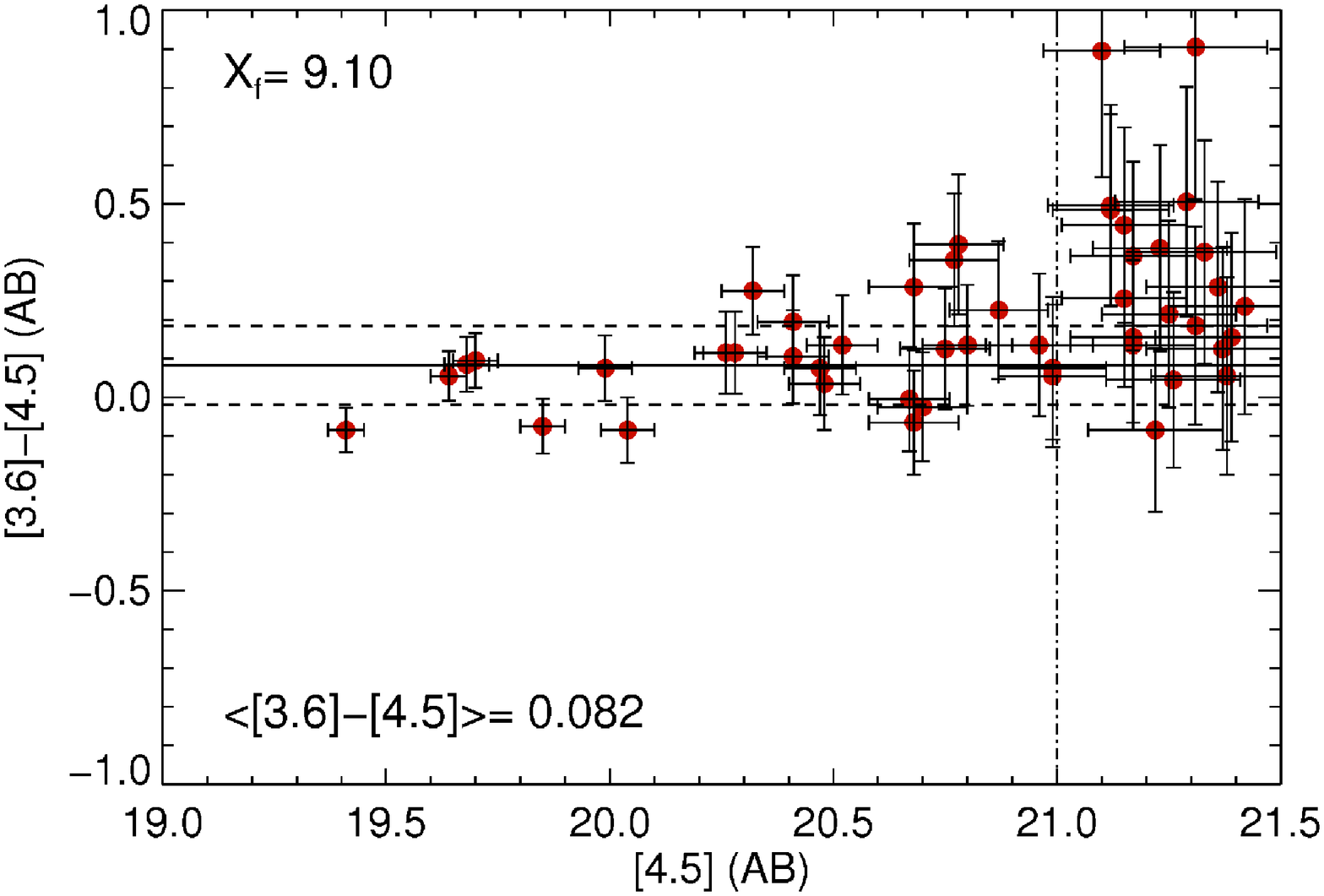}
\includegraphics[scale=.23,angle=-90]{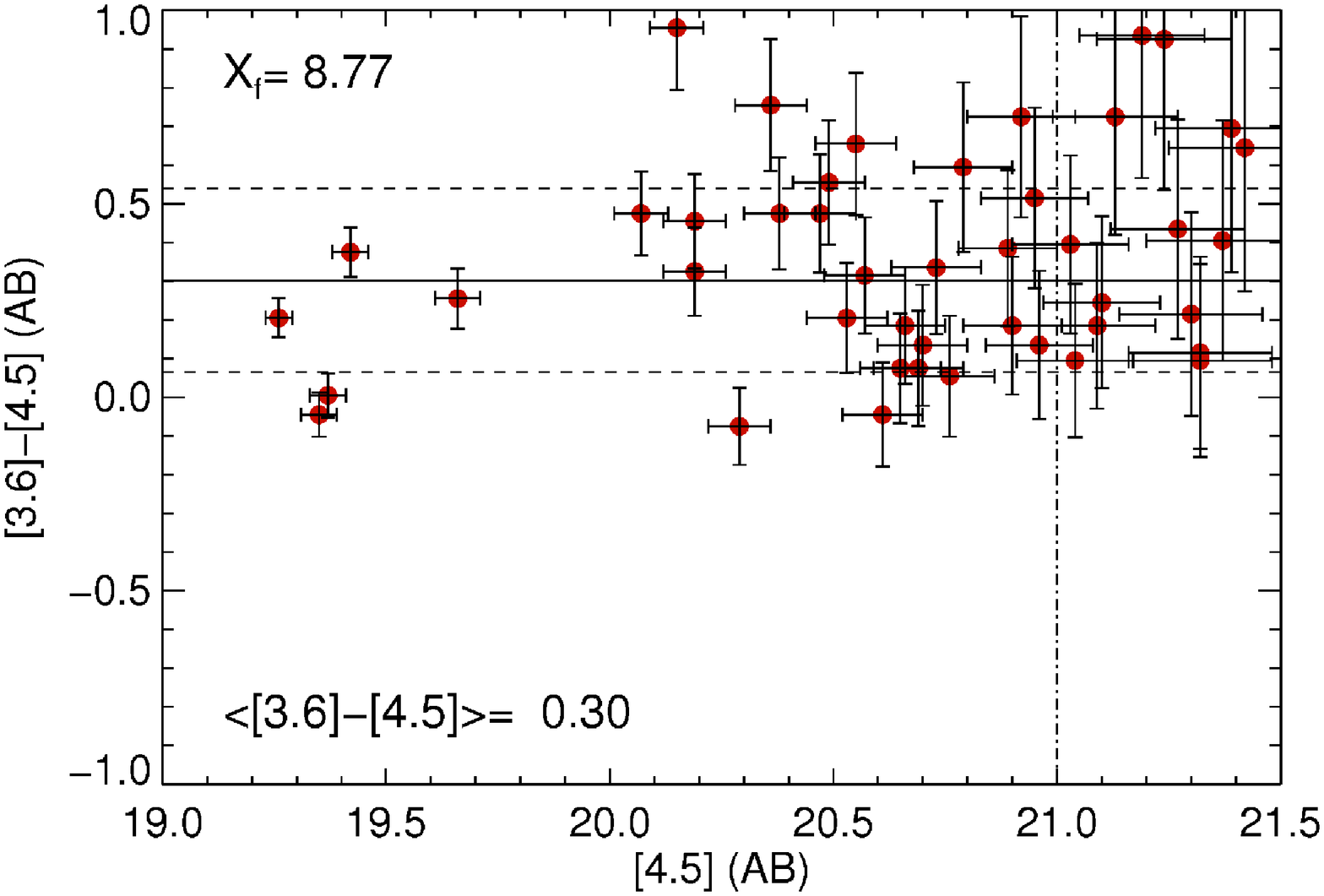}
\includegraphics[scale=.23,angle=-90]{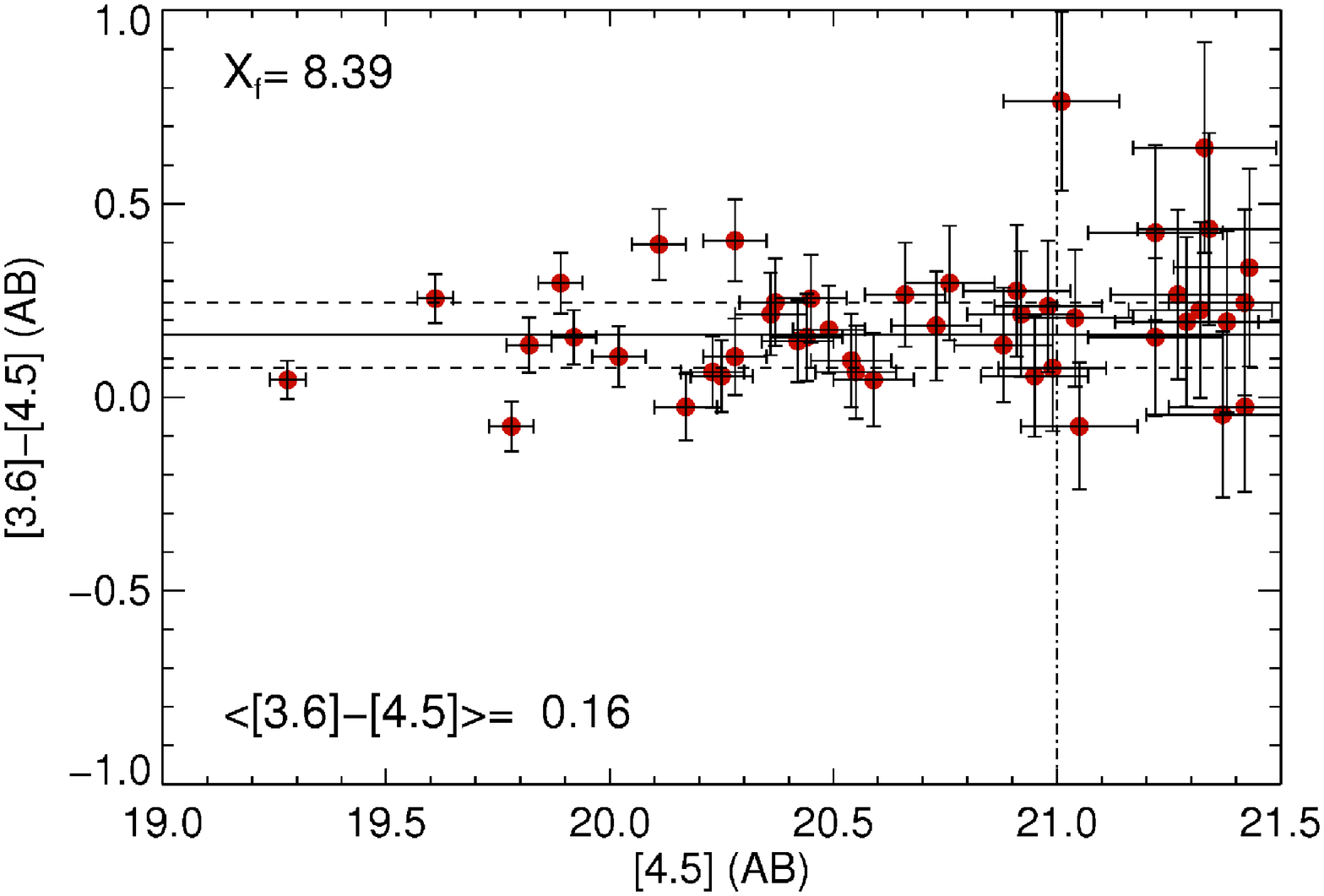}
\includegraphics[scale=.23,angle=-90]{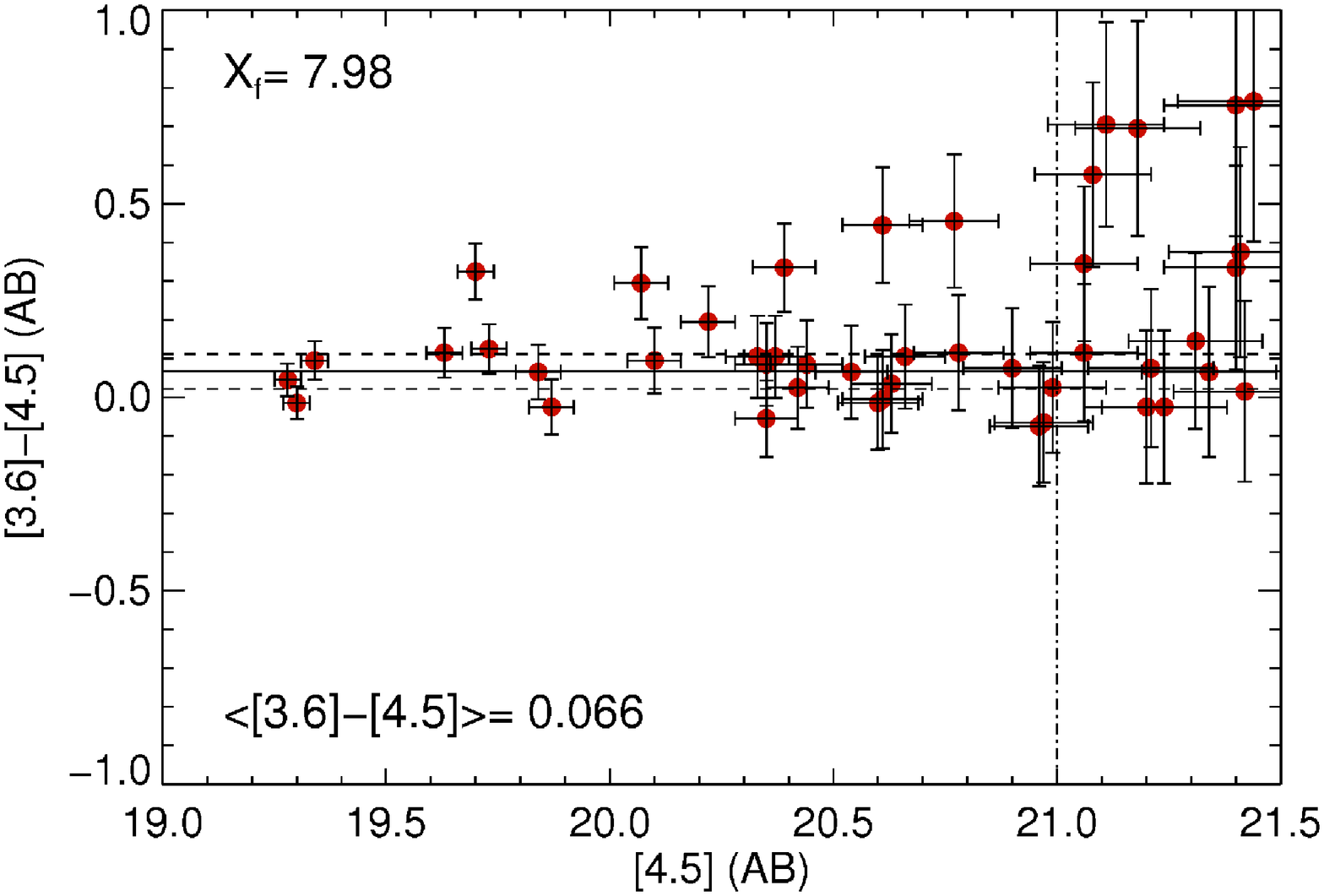}
\includegraphics[scale=.23,angle=-90]{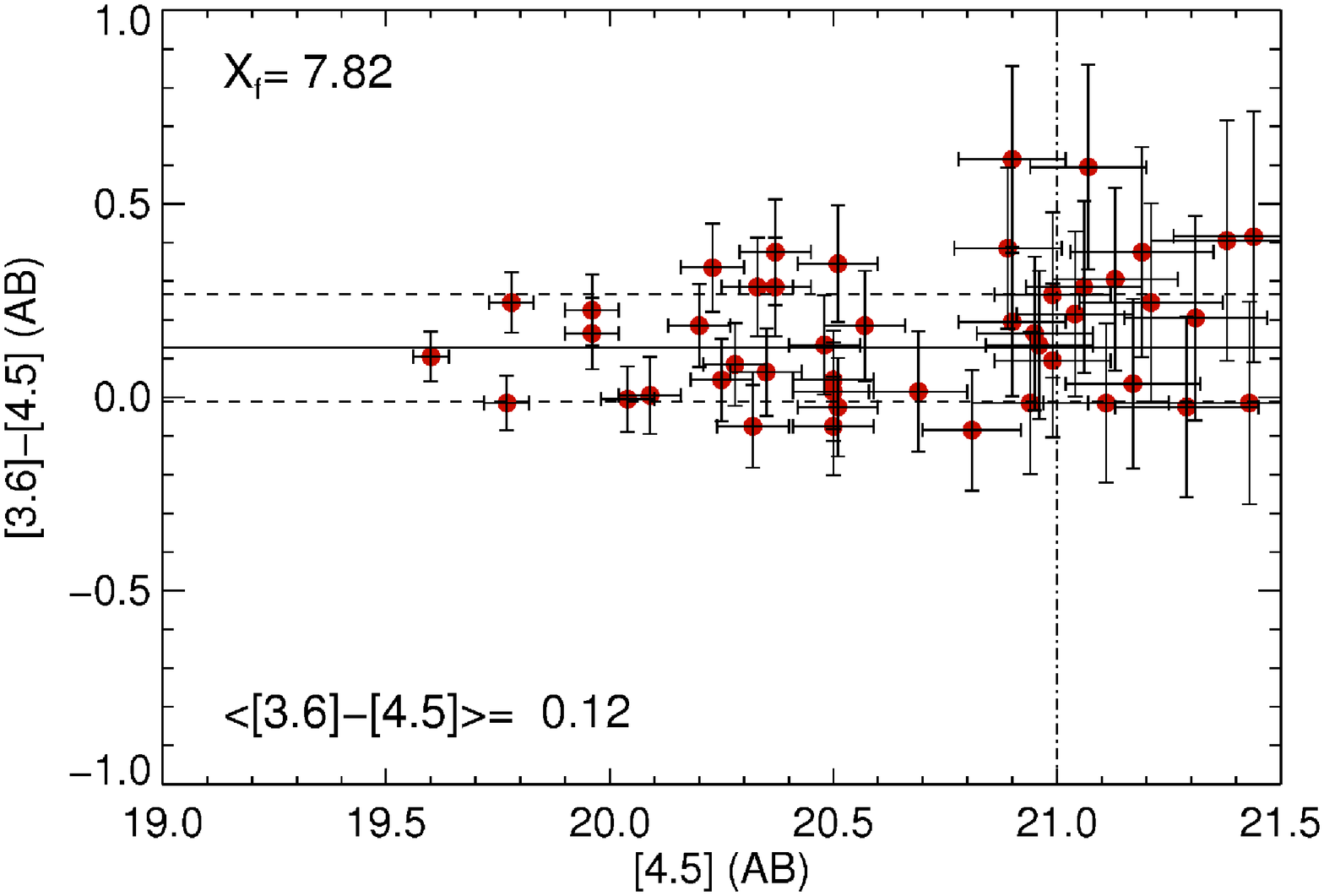}
\includegraphics[scale=.23,angle=-90]{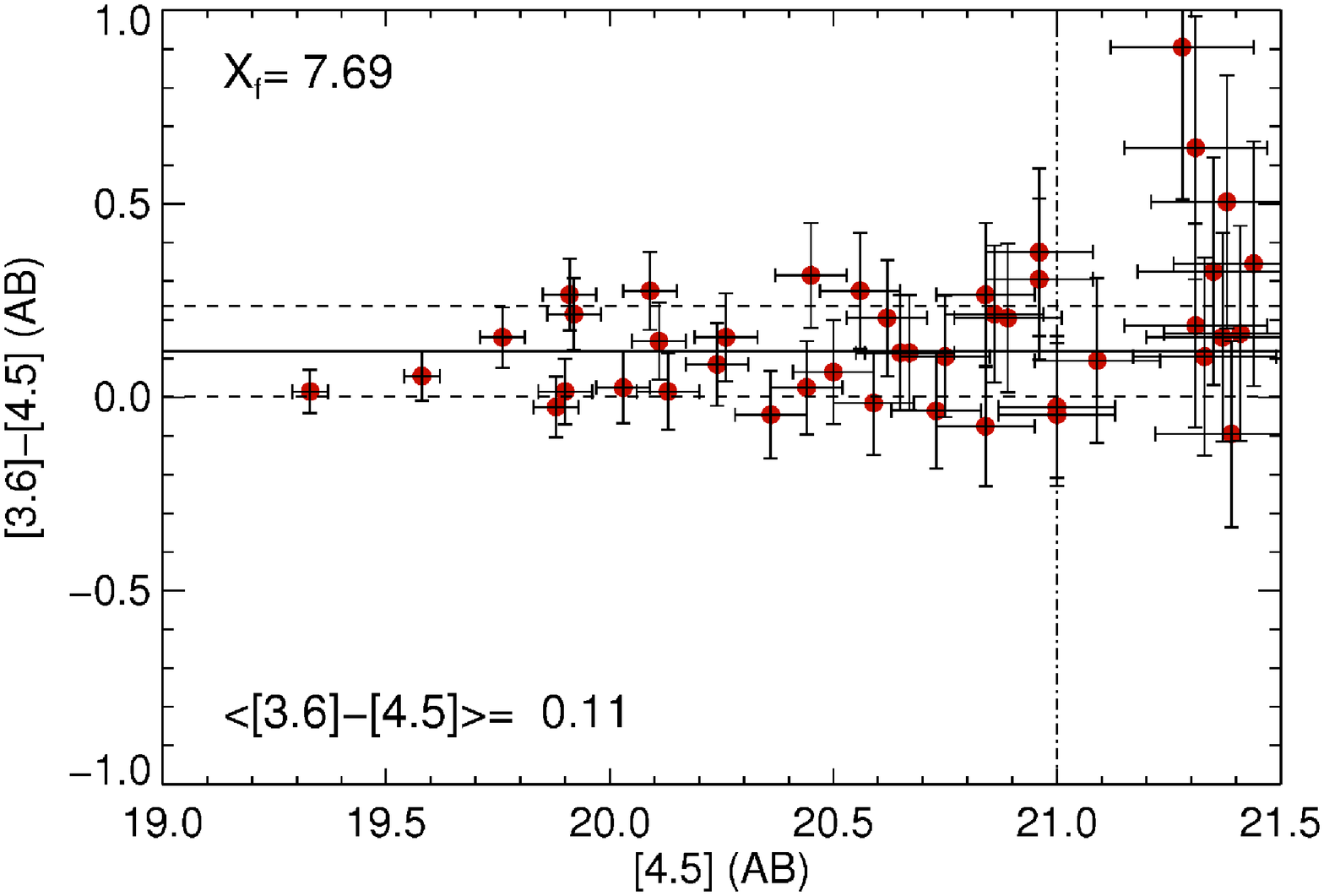}
\includegraphics[scale=.23,angle=-90]{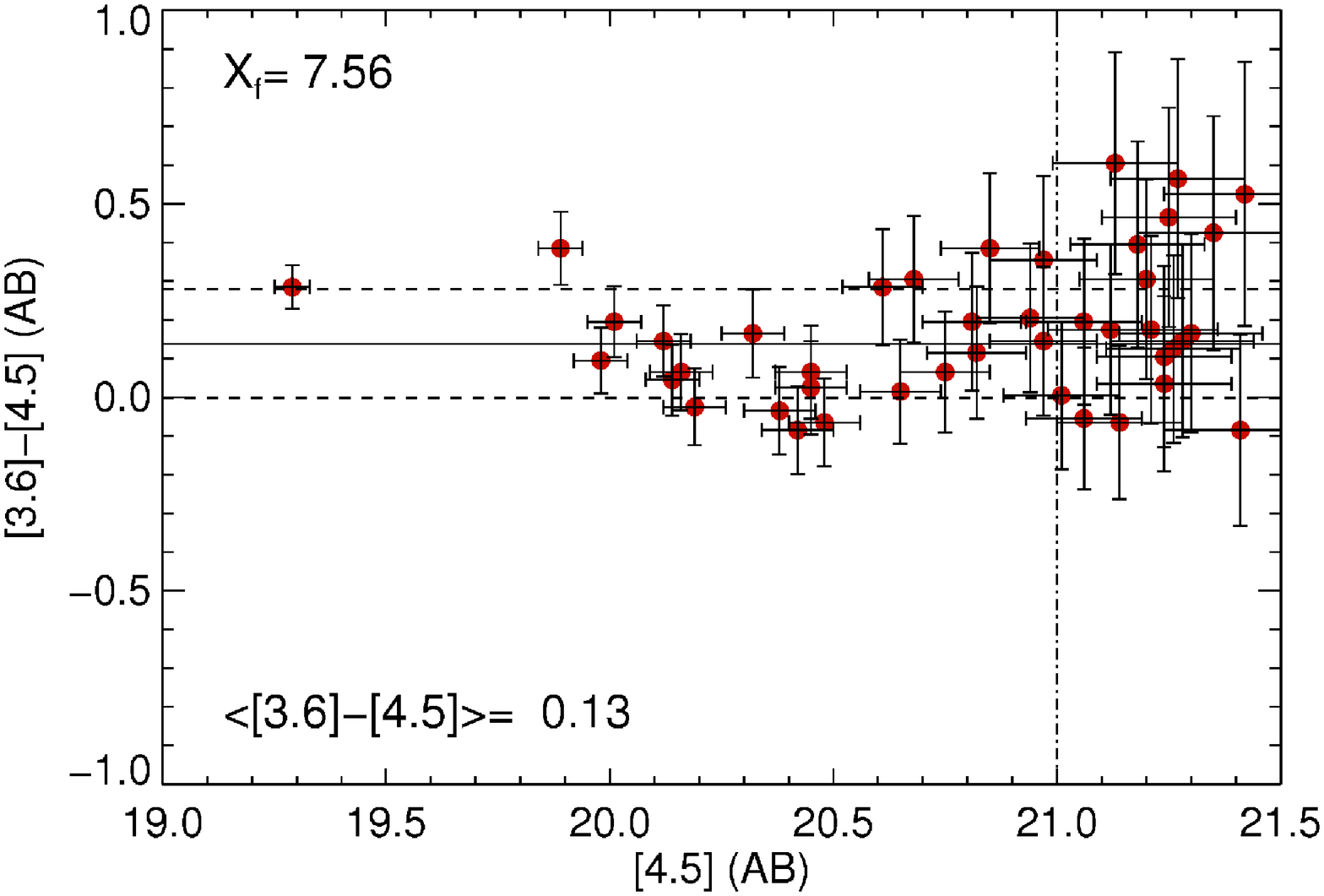}
\includegraphics[scale=.23,angle=-90]{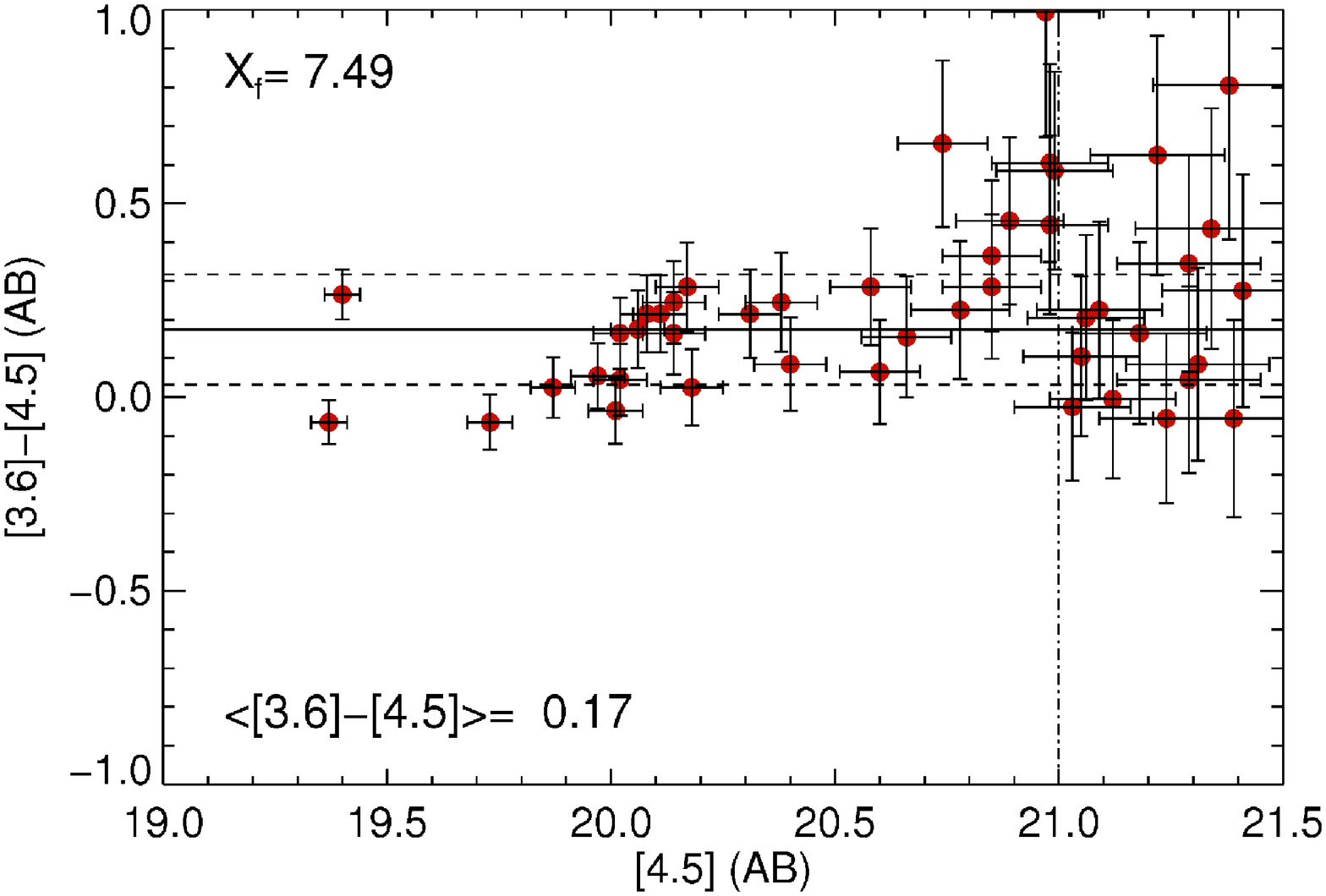}
\includegraphics[scale=.23,angle=-90]{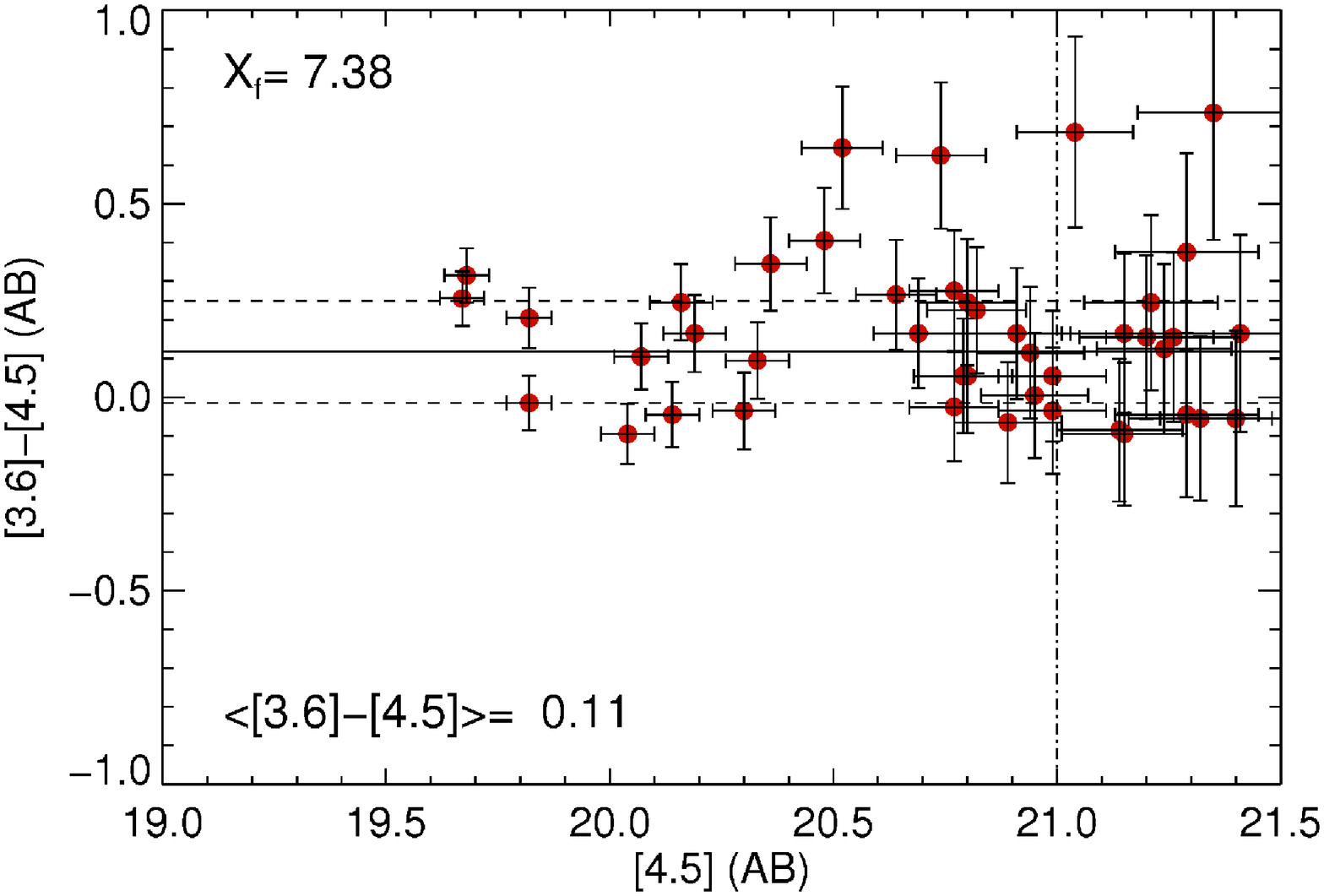}
\includegraphics[scale=.23,angle=-90]{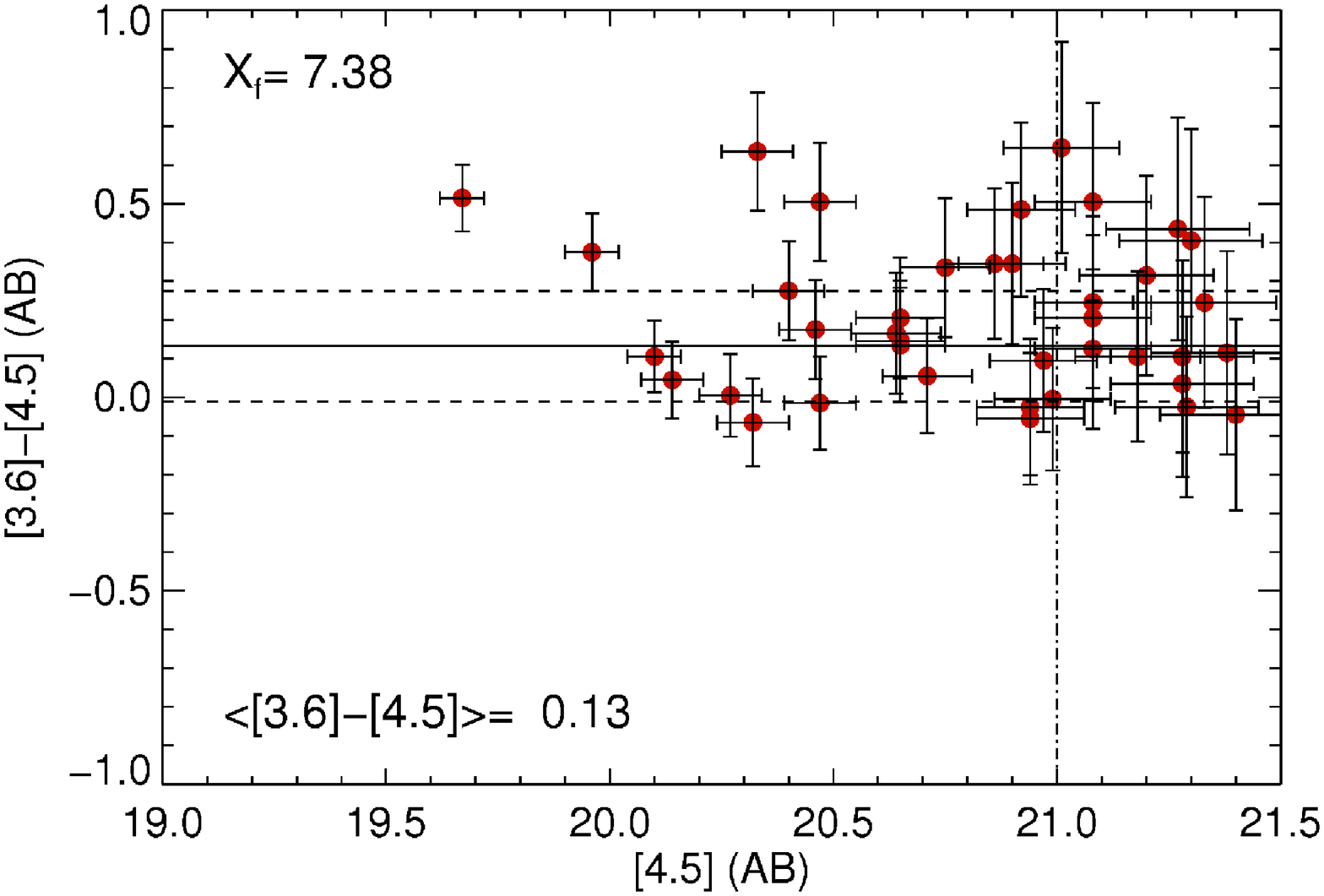}
\caption{$[3.6]-[4.5]$ color vs. $[3.6]$ magnitude for the 10 most significant high-redshift candidate galaxy clusters in the SSDF survey. The solid lines indicate the median $[3.6]-[4.5]$ color for member galaxies with $[4.5]< 21$. The dashed lines indicate $\pm \sigma$, the standard deviation of the color distribution.}
\label{clus_colmags}
\end{figure*}

In Table \ref{table_candidates}  we present the list of the top-10 most significant cluster candidates in the SSDF field. Figures \ref{clus_images} and \ref{clus_colmags} present their images and color-magnitude diagrams.

\begin{table}
  \caption{\label{table_candidates}  Top 10 Most Significant SSDF $z > 1.3$ galaxy cluster candidates} 
\label{table:1} 
\centering
\begin{tabular}{l c c c}     % 5columns
\hline\hline
\\
%ID &	RA &	DEC	 &  Excess Number &	$X_{f}$              \\
SSDF ID &	R.A. &	Dec.	 &   Significance              \\
  & (deg., J2000) & (deg., J2000) & ($X_{f}$) \\
\hline
\hline
%    250    &   354.786    &  -55.5308     &   34    &   9.1062   \\
%     355    &      0.049728     & -55.6782     &   32    &   8.7730     \\
%     247    &   355.235   & -54.0649   &    31    &   8.3905    \\
%     122     &  349.072     & -53.2995   &    29    &   7.9822     \\
%     269     &  359.440     & -58.0069     &   29     &  7.8274   \\
%     157    &   352.601    &  -50.6244    &    28     &  7.6940    \\
%     114    &   348.113    &  -52.8975    &   28     &  7.5600     \\
%     106   &    350.531    & -51.9481     &   28     &  7.4986     \\
%     131     &  349.301  &  -56.7126    &   27    &   7.3873     \\
%     155    &   353.632     & -50.8836   &    27    &   7.3853    \\
SSDF-CLJ2339-5531 &   354.786    &  $-$55.5308     &  9.10   \\%&  9.11   \\
SSDF-CLJ0000-5540 &  0.04973     & $-$55.6782     &  8.77     \\%&  8.77     \\
SSDF-CLJ2340-5403 &   355.235   & $-$54.0649   &  8.39    \\%&  8.39    \\
SSDF-CLJ2316-5317 &  349.072     & $-$53.2995   &  7.98     \\%&  7.98     \\
SSDF-CLJ2357-5800 &  359.440     & $-$58.0069     &  7.82   \\%&  7.83   \\
SSDF-CLJ2330-5037 &   352.601    &  $-$50.6244    & 7.69    \\%& 7.69    \\
SSDF-CLJ2312-5253 &   348.113    &  $-$52.8975    &  7.56     \\%&  7.56     \\
SSDF-CLJ2322-5156 &    350.531    & $-$51.9481     &  7.49     \\%&  7.50     \\
SSDF-CLJ2317-5642 &  349.301  &  $-$56.7126    &  7.38     \\%&  7.39     \\
SSDF-CLJ2334-5053 &   353.632     & $-$50.8836   &  7.38    \\%7.38    \\
\hline
\end{tabular}
\end{table}

\begin{figure*}
\epsscale{0.8}
\includegraphics[scale=0.23,angle=-90]{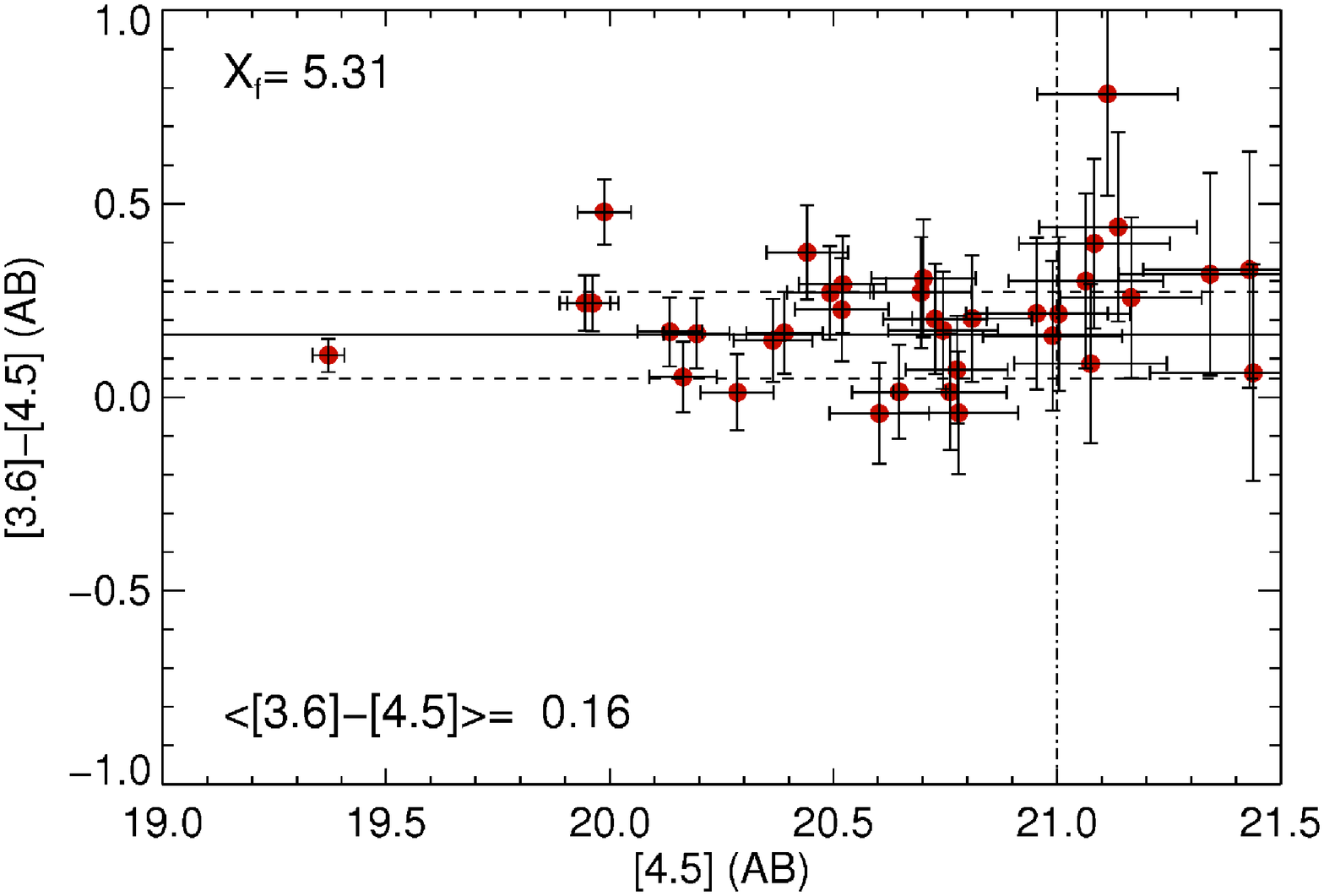}
\includegraphics[scale=0.23,angle=-90]{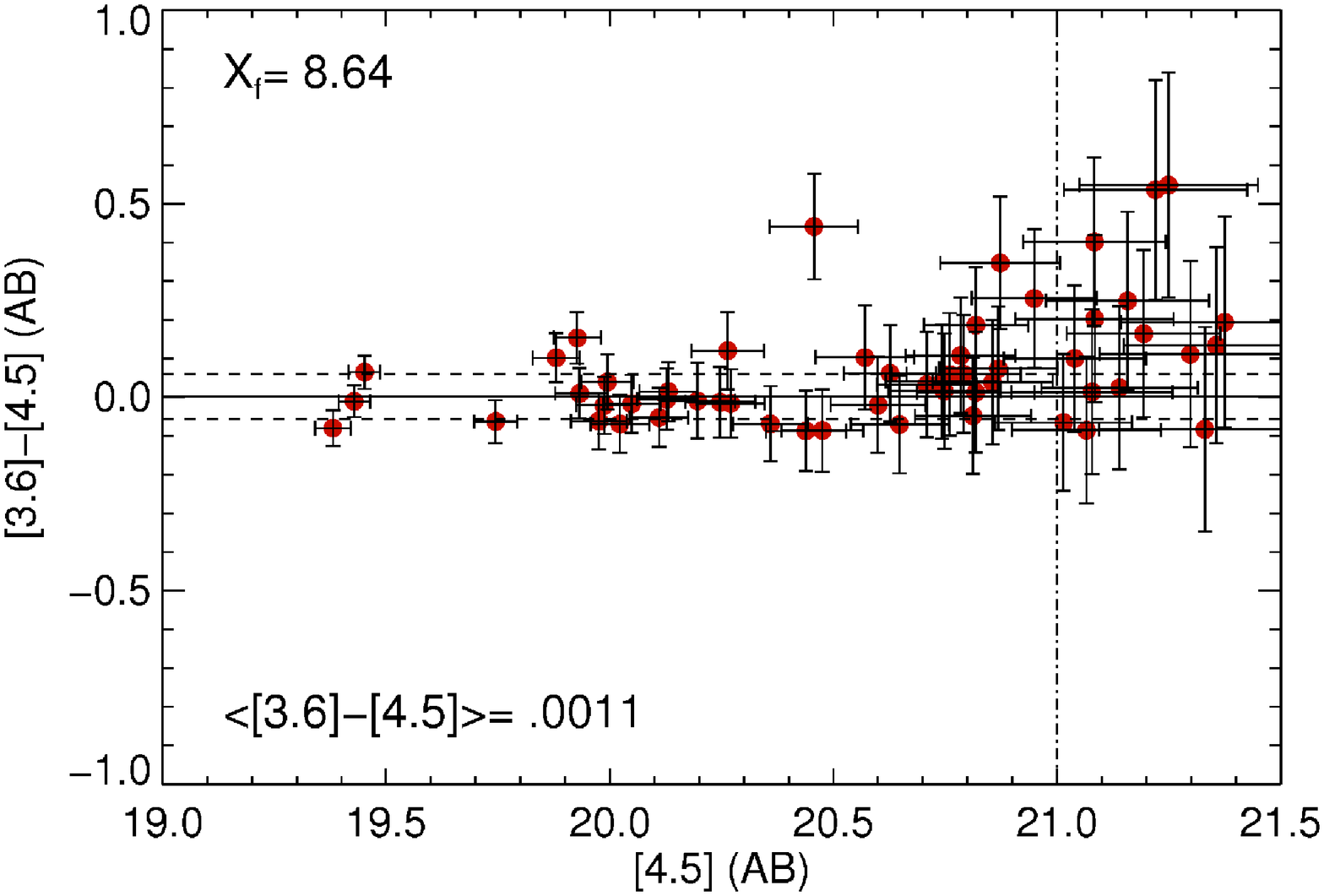}
\includegraphics[scale=0.23,angle=-90]{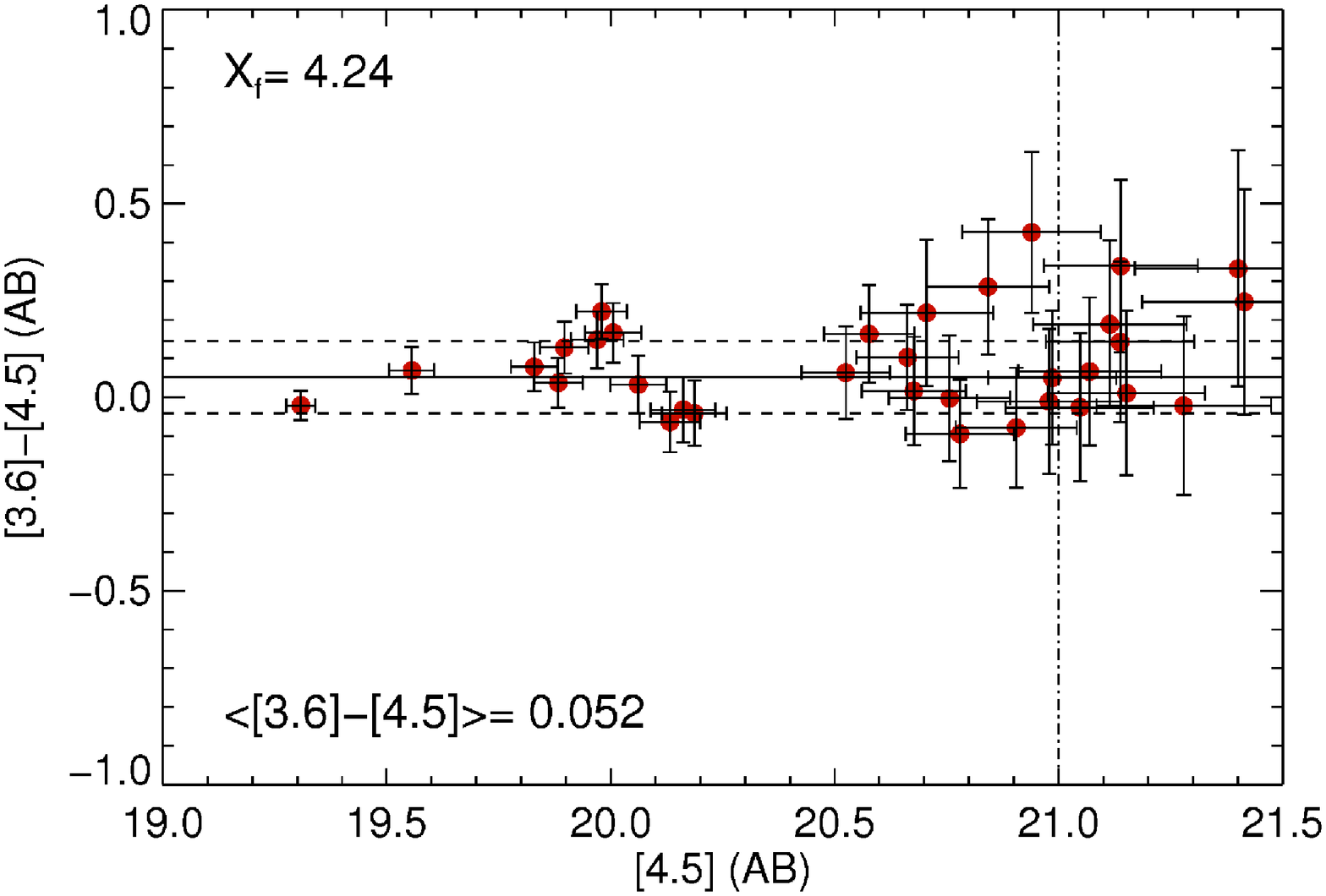}
\caption{$[3.6]-[4.5]$ color vs. $[4.5]$ magnitude for the IDCS J1426.5+3508 cluster confirmed at $z=1.75$ (left panel), as found by our algorithm with one the highest detection significance ($X_{f}=5.31$) when applied to comparable-depth observations from the ISCS. Color-magnitude diagrams are also shown for ISCS~J1438.1+3414 at $z = 1.41$  ($X_{f}=8.64$, middle panel) and ISCS~J1432.4+3250
at $z = 1.49$  ($X_{f}=4.24$, right panel), the two confirmed ISCS clusters with X-ray emission also found by our algorithm at high $X_{f}$. Lines and symbols are similar to those adopted for Fig. 5.}
\label{Obama}
\end{figure*}

\subsection{Purity of the SSDF Cluster Candidates}

For a large sample of candidate clusters to be useful for galaxy evolution and cosmological studies, it is important to determine its purity, $f_{{\rm pure}}$ as a function of the detection significance $X_{f}$,  defined as

\begin{equation}
f_{{\rm pure}} (X_{f})= \frac{N_{{\rm real}}}{N_{{\rm tot}}} = 1 - \frac{N_{{\rm false}}}{N_{{\rm tot}}},\label{purity}
\end{equation}

\noindent where $N_{{\rm tot}}$ is the total number of clusters candidates above the detection threshold $X_{f}$,  $N_{{\rm real}}$ is the number of candidates corresponding to real clusters and  $N_{{\rm false}}$ is the number of false detections. To this aim, we run our cluster finding algorithm on comparable-depth observations from the IRAC Shallow Cluster Survey (ISCS) \citep{Eisenhardt04}.

The ISCS is a wide-field IR-selected galaxy cluster survey carried out using 90s {\it Spitzer}/IRAC imaging of the 8.5 deg$^{2}$  Bo{\"o}tes field of the NOAO Deep, Wide-Field Survey \citep[NDWFS,][]{Jannuzi99}. SuperCOSMOS data are available in Bo{\"o}tes of comparable depth to the data available in the SSDF field. 
Over the past decade the Bo{\"o}tes cluster candidates have been the target of extensive ground- and space-based spectroscopic and photometric campaigns \citep[e.g., ][]{Eisenhardt08, Ashby09, Stern10, Brodwin11, Stanford12, Zeimann12, Gonzalez12, Brodwin13}. 
Using a wavelet search algorithm, operating on photometric-redshift probability distribution functions, described in \citet{Brodwin06}, more than 100 rich cluster candidates at $z>1$ were identified. To date  18 of these have been spectroscopically confirmed out to $z=1.9$.
Note that very accurate photometric redshifts measurements are based on data from the {\it Spitzer} Deep, Wide-Field Survey \citep[SDWFS;][]{Ashby09} (4 $\times$ deeper than the IRAC data used here), deep optical imaging in $B_{W}, R, I$ bands from the NDWFS, and NIR photometry from the FLAMEX survey \citep{Elston06} in the $J$- and $K_{s}$-bands. Note, however, that the ISCS is on-going, with many cluster candidates still awaiting confirmation.  In the absence of a complete, spectroscopically confirmed catalog of high-redshift clusters, we deem the ISCS the best available survey to test our algorithm as it contains the largest sample ($>10$) of spectroscopically confirmed clusters at $z>1.3$.

We find the purity of our sample,  $f_{{\rm pure}}$, to be a monotonic function of $X_{f}$  reaching $f_{{\rm pure}} = 0.8$ for $X_{f_{\rm min}}=5.2$. 
At this very high detection significance, our algorithm identifies 14 candidate clusters at $z > 1.3$ in the Bo\"otes field, of which three are spectroscopically confirmed at redshifts of $z = 1.37$ \citep{Brodwin13, Zeimann13}, $z = 1.41$ \citep{Stanford05, Brodwin11} and $z = 1.75$ \citep{Stanford12}, and an additional eight have accurate photometric redshifts $1.3 < z_{\rm phot} < 2.3$.
In the left panel of Fig.~\ref{Obama} we show the $[3.6]-[4.5]$ color vs. $[3.6]$ magnitude for the cluster  IDCS J1426.5+3508 at $z=1.75$ as found by our algorithm with $X_{f}=5.3$.
This cluster, spectroscopically confirmed with {\it HST}/WFC3 grism
observations \citep{Stanford12}, was detected in both archival 8.3~ks
{\it Chandra} imaging of the field  \citep{Stanford12} as well as follow-up
SZ observations with the CARMA array \citep{Brodwin12}.  The 
cluster also has a giant arc in {\it HST} imaging, implying it 
is a lensing cluster; this is particularly surprising given the 
cluster redshift and the small area of sky surveyed \citep{Gonzalez12}. 
At the time of its discovery, IDCS~J1426.5+3508 was the highest redshift
cluster for which the SZ effect had been measured.  
One of the other confirmed clusters, ISCS~J1438.1+3414 at $z = 1.41$, was the most
distant cluster known at the time of its discovery \citep{Stanford05}.
This inspired a deep, 145.0~ks observation with {\it Chandra} which
detected the cluster \citep{Brodwin11}. This cluster is found by our algorithm with the highest
significance ($X_{f} = 8.64$, see middle panel of Fig.~\ref{Obama}) in the Bo\"otes field.
The only other confirmed high-redshift Bo\"otes cluster with X-ray emission is ISCS~J1432.4+3250
at $z = 1.49$, which is also identified as an overdense region by
our algorithm, though at a level slightly below the very conservative threshold adopted
here ($X_{f} = 4.2$, see right panel of Fig.~\ref{Obama}). The red-sequence color and scatter for the known spectroscopically confirmed clusters in  Bo\"otes is very similar to the one shown by the SSDF candidates. To summarize, the conservative cut in detection significance we have derived from this analysis yields many previously identified clusters in the  Bo\"otes survey out to $z=1.8$, lending confidence in the effectiveness of our cluster finding algorithm when applied to similar-depth data in the SSDF.

\section{Clustering of High Redshift Clusters}

Clusters of galaxies reside in the largest dark matter  halos, representing higher density peaks of the mass distribution as redshift increases \citep{Springel05}. Thus the number density of galaxy clusters with redshift is sensitive to the cosmic matter density, $\Omega_{\rm m}$, and its evolution with redshift provides constraints on cosmological parameters \citep[e.g.,][]{Kitayama96, Wang98}. 

Several observational studies have previously demonstrated that cluster samples show strong clustering \citep[e.g.,][]{Bahcall88, Huchra90, Postman92, Borgani99, Gonzalez02, Bahcall03, Brodwin07, Papovich08} as measured by their autocorrelation function \citep{Peebles80}. 

High-resolution simulations predict that the cluster correlation function strength increases with redshift for a given mass limit. That is, high-redshift clusters are more strongly clustered, on a comoving scale, than low-redshift clusters of the same mass \citep{Bahcall03}. Moreover, a relative constancy of the cluster correlation function strength for the most massive clusters at every epoch is predicted: the $N$ most massive clusters at one epoch should have similar clustering to the $N$ most massive clusters at a later epoch \citep{Younger05}. 

Currently, both the measured correlation function scale lengths and number densities of several galaxy cluster samples support predictions from standard cold dark matter models ($\Lambda$CDM)  \citep[e.g., ][]{Abadi98, Croft97, Peacock92, Bahcall93, Collins00, Bahcall03, Brodwin07, Papovich08, Vikhlinin09, Hasselfield13}. Next, we measure these quantities for our cluster sample in order to check our consistency with previous observational studies and model predictions.

\subsection{Comoving number density}

\begin{figure}[t!]
\includegraphics[trim=08mm 25mm 15mm 25mm,clip=True,width=\columnwidth]{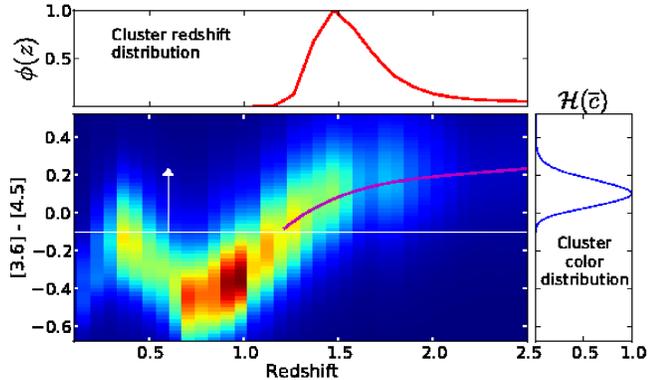}
\caption{{\bf Main}: $[3.6]-[4.5]$ color versus redshift distribution of the COSMOS galaxy catalog. In the SSDF we select galaxies with $[3.6]-[4.5]>-0.1$, marked as the horizontal white line. The magenta curve follows the peak of the distribution, where cluster colors will most likely be centered at. {\bf Right}: Distribution of mean colors in our cluster sample. {\bf Top}: Redshift distribution of our cluster sample. It is derived by matching the mean color distribution to the magenta path.}
\label{fig_dndz}
\end{figure}

%\subsection{Redshift distribution}
In order to derive the comoving number density of our cluster sample, we first  need to infer the expected redshift distribution $\phi(z)$ of our sample. To this aim, we employ the COSMOS/UltraVista photometric redshift catalog from \citet{Muzzin13}. This catalog includes 3.6$\mu$m and 4.5$\mu$m photometry reaching 2 magnitudes deeper than the SSDF. The central panel in Figure \ref{fig_dndz} shows the smoothed [3.6]-[4.5] color versus redshift distribution of the COSMOS catalog; 
details on this filtering procedure can be found in \citet{Martinez14}. The white line marks the  $[3.6]-[4.5]>-0.1$ selection, and the magenta curve follows the peak of the full galaxy distribution. We parametrize this curve with redshift and denote it as $\mathcal{S}(z)$. At a given redshift, we expect the color distributions of  cluster and field galaxies to have the same centroid (given by $\mathcal{S}(z)$), but with a smaller scatter for the clusters given their more homogeneous star formation histories and faster evolution \citep{Rettura10}. As pointed out in \S 3.1, a significant contamination by $z\sim0.3$ occurs with a simple IRAC color cut. As we remove this contamination with optical and $[4.5]$ magnitude cuts, in the following analysis we assume that all our clusters are  at $z>1$.

To analyze the colors of our cluster candidates, we consider the 10 brightest galaxies in $[4.5]$ from each of them (the total number of member candidates is $\sim40$ per cluster). The goal of this selection is two-fold: it removes faint galaxies which are more likely to be contaminants due to photometric errors, and keeps the most massive galaxies which are more likely to share a similar star formation history. The typical scatter in color in clusters after applying this selection is $\sim0.12$\,mag. In comparison, field galaxies at a given redshift have a much larger color scatter of $\sim0.25$\,mag, confirming our expectation that the cluster members are a more homogeneous population. 

At this point, we have established that our clusters are likely to follow $\mathcal{S}(z)$. In order to derive the cluster redshift distribution, we populate this curve by matching the distribution of mean colors from our full cluster sample. For a given cluster, we define the probability density function of having a mean color $\bar{c}$ as a normal distribution centered at $\bar{c}$ and standard deviation equal to the standard error of the colors from the 10 brightest cluster members. Then, the full distribution of mean colors, $\mathcal{H}(\bar{c})$, is given by the sum of the probability density functions from all clusters. This is shown in the right panel of Figure \ref{fig_dndz}. The redshift distribution is then derived as

\begin{equation}
\phi(z) dz = \mathcal{H}(\mathcal{S}(z)) \frac{d\mathcal{S}(z)}{dz} dz,\label{eq_color}
\end{equation}
which is displayed in the upper panel of the same figure.

We can calculate the spatial number density of the cluster sample at the pivot redshift $z_p \equiv 1.5$ by combining the redshift distribution $\phi(z)$, the SSDF survey area and the number of cluster candidates, $N_\mathrm{obs}$. The number of clusters within $z_p \pm \delta z/2$ can be written as

\begin{equation}
\Delta N = N_\mathrm{obs} \frac{\phi(z_p)}{\int \phi(z^\prime)dz^\prime} \delta z.
\end{equation}
The sampled volume reads as
\begin{equation}
\Delta V  = \frac{dV(z_p)}{dz}\delta z = \frac{c \, \Omega \, \chi^2(z_p)}{H(z_p)}\delta z,
\end{equation}
where $\chi(z)$ is the comoving radial distance, $H(z)$ is the Hubble function, $c$ is the speed of light and $\Omega=0.0271$ steradians is the solid angle subtended by the SSDF survey. Hence, the number density for our SSDF cluster sample, $n_{\rm {c}}$, at $z=1.5$ is 
\begin{equation}
n_{\mathrm {c}} = \frac{\Delta N}{\Delta V}  = (7.6 \pm 0.6) \times 10^{-7} h^{3} \mathrm{Mpc}^{-3} . \label{eq_nc_obs}
\end{equation}

\subsection{Halo model}
Our infrared selection of galaxies and the ranking system we adopt for our cluster search algorithm is expected to produce a nearly mass-limited cluster sample at $z>1.3$. Thus, we assume that our cluster sample is comprised of most dark matter haloes above some characteristic minimum mass $M_\mathrm{min}$, following the halo occupation framework \citep{ma&fry00,seljak00,peacock00,scoccimarro01, cooray02, berlind02,berlind03, kravtsov04,zheng05}. We model the probability of a halo of mass $M$ to be part of the cluster sample as:

\begin{equation}
 N_c(M) = \left\{
\begin{array}{l l}
    1 & \quad \text{if $M \geq M_\mathrm{min}$}\\
    0 & \quad \text{if $M < M_\mathrm{min}$}
\end{array} \right\}
\label{e_nc}
\end{equation}
Then, the comoving number density of clusters is determined as:

\begin{equation}
n_c = \int\limits_{M_\mathrm{low}}^{M_\mathrm{high}}dM \frac{dn}{dM}(M)  N_c(M|M_\mathrm{min}) \label{eq_nc_model}
\end{equation}
where $\frac{dn}{dM}(M)$ is the halo mass function from \citet{tinker10} and the integration limits are $M_\mathrm{low}=10^{11} M_\odot$ and $M_\mathrm{high}=10^{16} M_\odot$. In addition, we can calculate the mean mass of the cluster sample as an average over the occupied halos:

\begin{equation}
M_\mathrm{mean} =\frac{1}{n_c}\int dM \frac{dn(M)}{dM}N_c(M|M_\mathrm{min})\,M. \label{e_meff}
\end{equation}

\subsection{Angular Correlation Function}

An additional observable that links the distribution of dark matter halos and our clusters is the measurement of their clustering. In particular, we focus on the modeling of the two-point spatial correlation function (SCF, formally $\xi(r)$, Peebles 1980), which represents the excess probability of finding two objects at a separation $r$ with respect to the case of a randomly distributed sample. The formation of dark matter halos of a given mass $M$ is directly correlated with the dark matter overdensity at their location \citep{kaiser84,fry93, mo&white96}. 
This translates into a scaling between the dark matter and halo SCFs, which is called the halo bias $b_h$:   

\begin{equation}
\xi_h(M,r,z) = \xi_{dm}(r,z) \, b_h^2(M,z).   \label{eq_bias}\\
\end{equation}
Our definition of halo mass is that enclosed by a sphere with a density 200 times larger than the critical density of the Universe. We obtain the dark matter correlation function from the CAMB software \citep{lewis02}, and the halo bias from \citet{sheth01}, using the updated parameters from \citet{tinker05}.  The bias of the clusters with respect to dark matter then reads

\begin{equation}
b_c = \frac{1}{n_c}\int dM \frac{dn(M)}{dM}N_c(M|M_\mathrm{\rm min})b_h(M). \label{e_bg}
\end{equation}

Since the observed configuration space of our cluster sample is the celestial sphere, we measure its clustering via the angular correlation function (ACF, formally $\omega(\theta)$). The ACF can be considered as the radial projection of the SCF, which can be computed with the knowledge of the sample's redshift distribution \citep{limber53, phillips78}:

\begin{equation}
\omega(\theta) = \frac{2}{c} \int_0^\infty dz H(z) \phi^2(z) \int_0^\infty dy \,\, \xi_c (r=\sqrt{y^2+D_c^2(z)\theta^2}), \label{eq_limber}
\end{equation}
where $ \phi(z)$ is the normalized redshift distribution, $D_c(z)$ is the radial comoving distance, $c$ is the speed of light and $\theta$ is the angular separation given in radians. We measure $\omega(\theta)$ with the estimator presented in \citet{hamilton93}, which counts the number of galaxy pairs with respect to those of a random sample distributed in the same geometry:

\begin{equation}
\hat{\omega}(\theta)=\frac{\mathrm{RR}(\theta) \mathrm{GG}(\theta) }{\mathrm{(GR)}^2(\theta)} - 1, \label{eq_ham}
\end{equation}
where GG, GR and RR are total number of galaxy-galaxy, galaxy-random and random-random pairs separated by an angle $\theta$. There is no need to include a correction for the integral constraint \citep{Peebles80}, since it was shown in \citet{Martinez14} that it is negligible for this very wide-field survey geometry. We estimate $\omega(\theta)$ errors using jackknife resampling. For this, the entire sample is divided into $N_\mathrm{jack}=32$ spatial regions of equal size. Then, the correlation is run $N_\mathrm{jack}$ times, each one excluding one of those regions from the sample. The value of the estimator is the average $\bar{\omega}(\theta)$ of those iterations and the covariance between angular bins is given by 

\begin{equation}
C_{jk} = \frac{N-1}{N}\sum\limits_{i=0}^N \left[ \hat{\omega}_i(\theta_j)-\bar{\omega}(\theta_j) \right]\left[ \hat{\omega}_i(\theta_k)-\bar{\omega}(\theta_k) \right]. \label{eq_Cij}
\end{equation}
The observed $\omega(\theta)$ function is shown in Figure \ref{fig_acf}, where the error bars are derived from the diagonal elements of the covariance matrix.

\begin{table}
\begin{center}
\caption{\label{table_clustering} Best-fit results for the number density, the characteristic minumum mass, the mean mass, the bias and the angular correlation length of the SSDF cluster sample as found by the density and the clustering fits.}
  \begin{tabular}{ccc}
 & Density& Clustering \\
\hline \hline \\[-0.2cm]
$n_c$ ($10^{-7} h^{3} \mathrm{Mpc}^{-3}$) & $7.6 \pm 0.6  $ & $0.7^{+6.3}_{-0.6} $\\[0.2cm] \hline \\[-0.2cm]
$M_\mathrm{min}$ ($10^{14} h^{-1} M_\odot$)& $0.78\pm0.02$& $1.5^{+0.9}_{-0.7}$\\ [0.2cm] \hline\\[-0.2cm]
$M_\mathrm{mean}$ ($10^{14} h^{-1} M_\odot$)& $1.08\pm0.02$& $1.9^{+1.0}_{-0.8}$\\ [0.2cm] \hline\\[-0.2cm]
$b_c$ & $8.2\pm0.1$&$10.8\pm2.5$\\ [0.2cm] \hline\\[-0.2cm]
$r_0$ ($h^{-1} \mathrm{Mpc}$)& $25.9\pm0.4$& $ 32\pm7$ \\  [0.1cm]
\hline
\end{tabular}
\end{center}
\end{table}

\subsubsection{Results}
The halo model of the clustering described so far can be fully specified either by fixing $n_c$ through the observed cluster counts (Equation \ref{eq_nc_obs}), or by fixing $b_c$ through the amplitude of the observed clustering. We refer to these two approaches as the density and clustering fits, respectively. Thus, comparing these sets of results is an excellent way to test the consistency of the model and observations. Our fits are based on the maximization of the likelihood of the model given the data, $\mathcal{L}(${\footnotesize mod}$|${\footnotesize data}$)=e^{-\chi^2/2}$. For the clustering fit, this is specified by:
\begin{equation}
\chi^2 = \sum\limits_{i=0}^K \sum\limits_{j=0}^K \left[ \omega_m(\theta_j)-\bar{\omega}(\theta_j) \right] C_{ij}^{-1}  \left[ \omega_m (\theta_k)-\bar{\omega}(\theta_k) \right]. \label{chi2}
\end{equation}
Here, $\omega_m$ and $\bar{\omega}$ are the predicted and observed ACFs (Equations \ref{eq_limber} and \ref{eq_ham}, respectively), $C_{ij}$ is the covariance matrix from Equation \ref{eq_Cij} and $K=4$ is the number of angular bins. 

For the density fit, 
\begin{equation}
\chi^2=(n_c^\mathrm{mod}-n_c^\mathrm{data})^2/\sigma^2_{n_c},  \label{chi2density}
\end{equation}

\noindent where the variance $\sigma^2_{n_c}$ is derived from Poisson statistics in the cluster counts. Results are shown in Table \ref{table_clustering}. 
We have included the calculation of $r_0$, which marks the distance scale length where $\xi_c(r_0)=1$, with $\xi_c = \xi_{dm}(z=z_p) b_c^2 $ for both methods. 

\begin{figure}[th!]
\includegraphics[scale=.47]{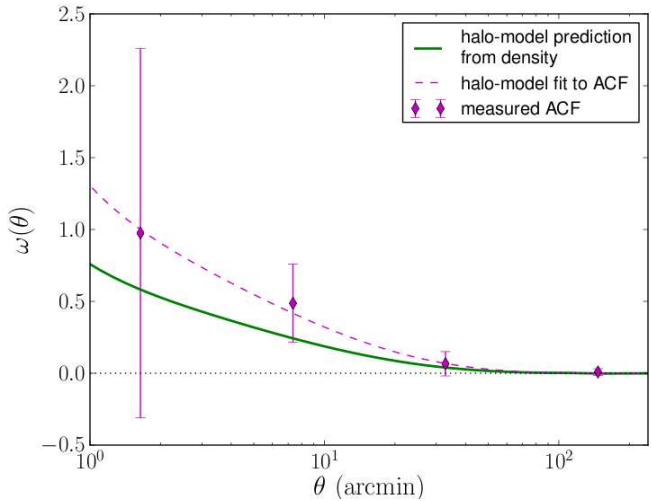}
\caption{\label{fig_acf} Observed clustering measured via the ACF (points), fitted $\omega(\theta)$ (dashed curve) and predicted $\omega(\theta)$ based on the observed number density (solid curve).  }
\end{figure}

Overall, there is a reasonable agreement between the density and clustering sets, differing by less than 2~$\sigma$. Throughout the paper, we make the conservative choice of adopting the clustering sets as fiducial values since they have the larger errors. The two angular correlation functions, $\omega(\theta)$,  one corresponding to the prediction based on the number density (solid curve) and one based on the fit of the measured ACF points (dashed curve) are displayed in Figure \ref{fig_acf} and found to be consistent within the errors.

The values of $n_{\rm c}$  and $r_0$ found for our sample of candidate clusters are consistent with those found by previous observational studies in the literature \citep[][see also Fig. 9 of Papovich 2008]{Abadi98, Collins00, Bahcall03, Brodwin07} and are well in agreement with predictions based on $\Lambda$CDM cosmological models \citep{Springel05}. This analysis lends further confidence in the effectiveness of our cluster finding algorithm.

\section{Summary}

We have identified a large sample of massive high-redshift galaxy cluster candidates of galaxies at $z>1.3$ over the 94 deg$^2$ {\it Spitzer} survey of the SPTpol field.  Our algorithm identifies the most significant overdensities of galaxies based upon their IRAC color ($[3.6]-[4.5]>-0.1$), their $4.5 \mu$m magnitude ($[4.5]>19.5$)  and requiring non-detection in the shallow SuperCOSMOS $I$-band data ($I > 20.45$).  We identify 279 distant cluster candidates using a $X_{f} \geq 5.2$ detection significance, for which we estimate a $\sim 80\%$ purity by running our algorithm on comparable-depth observations of the  {\it Spitzer} surveys of the Bo{\"o}tes field \citep{Eisenhardt04, Ashby09}, which has been the target of extensive ground- and space-based spectroscopic and photometric campaigns over the past decade.

We find that the SSDF cluster sample shows strong clustering.  From the angular correlation analysis, we find our sample has a comoving number density $n_c  = (0.7^{+6.3}_{-0.6}) \times 10^{-7} h^{3} \mathrm{Mpc}^{-3}$%$n_c  = (7.6 \pm 0.6 ) \times 10^{-7} h^{3} \mathrm{Mpc}^{-3}$
 and a spatial clustering correlation scale length $r_0 = (32 \pm 7) h^{-1} \rm{Mpc}$%$r_0 = (25.9 \pm 0.4) h^{-1} \rm{Mpc}$
. These values are consistent with previous observational studies and match expectations based on $\Lambda$CDM high-resolution simulations. 
The high-redshift cluster sample presented here  has a mean mass $M_{{\rm mean}} = 1.9^{+1.0}_{-0.8} \times 10^{14} h^{-1} M_{\odot}$.
Assuming these clusters grow according to predictions of $\Lambda$CDM \citep[e.g.,][]{Fakhouri10}, they will evolve into massive clusters ($> 5 \times 10^{14} h^{-1} M_{\odot}$) at $z=0.2$.

\smallskip
This study showcases the impact that large {\it Warm Spitzer} surveys can have on the identification of large samples of massive clusters of galaxies at very high redshifts in the upcoming years.
In particular, this sample has been selected in an area where deep observations for the SZ effect with the SPTpol camera are underway and part of this field has also {\it XMM-Newton} deep X-ray observations from the XXL Survey. These ancillary data will allow us to determine cluster masses for our sample, enabling systematic study of the cluster population in a crucial epoch for their assembly. 
%Furthermore, optical and infrared spectroscopic campaigns targeting individual candidate cluster galaxies with Magellan, Gemini-S and VLT telescopes will be crucial to assign cluster membership and will surely stimulate significant progress in the years to come.

\acknowledgments  A.R.  is  grateful to the SSDF Team for providing access to advanced data products and is thankful to Audrey Galametz, Dominika Wylezalek, Loredana Vetere and Roberto Assef for useful discussions and comments on this paper. This work is based on data obtained with the {\it Spitzer Space Telescope}, which is operated by the Jet Propulsion Lab (JPL), California Institute of Technology (Caltech), under a contract with NASA. Support was provided by NASA through contract number 1439211 issued by JPL/Caltech.  Lawrence Livermore National Laboratory is operated by Lawrence Livermore National Security, LLC, for the U.S. Department of Energy, National Nuclear Security Administration under Contract DE-AC52-07NA27344.

\email{aastex-help@aas.org}

%% To help institutions obtain information on the effectiveness of their
%% telescopes, the AAS Journals has created a group of keywords for telescope
%% facilities. A common set of keywords will make these types of searches
%% significantly easier and more accurate. In addition, they will also be
%% useful in linking papers together which utilize the same telescopes
%% within the framework of the National Virtual Observatory.
%% See the AASTeX Web site at http://www.journals.uchicago.edu/AAS/AASTeX
%% for information on obtaining the facility keywords.

%% After the acknowledgments section, use the following syntax and the
%% \facility{} macro to list the keywords of facilities used in the research
%% for the paper.  Each keyword will be checked against the master list during
%% copy editing.  Individual instruments or configurations can be provided 
%% in parentheses, after the keyword, but they will not be verified.

%{\it Facilities:} \facility{Nickel}, \facility{HST (STIS)}, \facility{CXO (ASIS)}.
{\it Facilities:} \facility{Spitzer}, \facility{UKST}

%% Appendix material should be preceded with a single \appendix command.
%% There should be a \section command for each appendix. Mark appendix
%% subsections with the same markup you use in the main body of the paper.

%% Each Appendix (indicated with \section) will be lettered A, B, C, etc.
%% The equation counter will reset when it encounters the \appendix
%% command and will number appendix equations (A1), (A2), etc.

\clearpage

\end{document}